\documentclass[]{aastex63}

\usepackage[caption=false]{subfig}

\newcommand{\lya}{Ly$\alpha$}

\received{June 7, 2022}
\accepted{November 15, 2022}
\submitjournal{The Astrophysical Journal}

\shorttitle{Galaxies in the Vicinity of \lya\ Nebulae}
\shortauthors{Wells, Prescott \& Finlator}

\begin{document}

\title{Brighter and More Massive Galaxies in the Vicinity of \lya\ Nebulae}

\author[0000-0003-2101-3540]{Natalie K. Wells}
\affiliation{Department of Astronomy, New Mexico State University, P.O. Box 30001, MSC 4500, Las Cruces, NM, 88003, USA}

\author[0000-0001-8302-0565]{Moire K. M. Prescott}
\affiliation{Department of Astronomy, New Mexico State University, P.O. Box 30001, MSC 4500, Las Cruces, NM, 88003, USA}

\author[0000-0002-0496-1656]{Kristian M. Finlator}
\affiliation{Department of Astronomy, New Mexico State University, P.O. Box 30001, MSC 4500, Las Cruces, NM, 88003, USA}

\begin{abstract}

It has been well established in the local universe that galaxy properties differ based on the large-scale environment in which they reside. As luminous Lyman-alpha (Ly$\alpha$) nebulae have been shown to trace overdense environments at z$\sim$2-3, comparing the properties of galaxies within \lya\ nebulae systems to those in the field can provide insight into how and when locally-observed trends between galaxy properties and environment emerged. Six \lya\ nebulae were discovered at z$\sim$2.3 in a blind search of the GOODS-S extragalactic field, a region also covered by the 3D-HST spectroscopic survey. Utilizing 3D-HST data, we identified 86 galaxies in the vicinity of these nebulae and used statistical tests to compare their physical properties to galaxies elsewhere in the field. Galaxies lying within 320 proper kpc of a \lya\ nebula are roughly half a magnitude brighter than those in the field, with higher stellar masses, higher star formation rates, and larger effective radii. Even when considering the effects of sample incompleteness, our study suggests that galaxies in overdensities at z$\sim$2.3 traced by \lya\ nebulae are being influenced by their environment. Furthermore, \lya\ nebula-associated galaxies lie on the same main sequence of star formation as field galaxies, but have a larger proportion of high-mass galaxies, consistent with the idea that galaxy evolution is accelerated in rich environments. Expanded surveys for \lya\ nebulae in other deep extragalactic fields and galaxy spectroscopic follow-up with JWST will better constrain the demographics of \lya\ nebula-associated galaxies.

\end{abstract}


\section{Introduction} 
\label{sec:intro}

In the local universe (z$<$1), a galaxy’s structure is strongly dependent on its surroundings: elliptical galaxies are found predominantly in clusters, while disk galaxies are more common in isolation. This ``morphology-density relation'' was first established by \cite{Oemler1974} and \cite{Dressler1980}, and its physical origins are still the subject of investigation. The star formation rate (SFR) is also linked to environment and morphology at low redshifts, with cluster ellipticals tending to be quiescent and red, and field spirals showing blue colors and substantial SFRs \citep[e.g.][]{Lewis2002, Gomez2003, Hogg2004, Kauffmann2004, Peng2010}.  

Study of the more distant universe gives us a window back in time to when these environmental differences in galaxy properties first emerged. In high redshift studies (z=2-4), galaxies in dense protocluster environments have widely been found to have higher stellar masses when compared to a control group \citep{Koyama2013, Hatch2011, Steidel2005, Cooke2014, Shimakawa2017, Ito2020}. Somewhat less agreement is found, however, when examining growth histories. For example, \cite{Koyama2013} and \cite{Cooke2014} found that protocluster and field galaxies follow the same SFR-stellar mass relation at z$\sim$2.2 and z$\sim$2.5, respectively, as did \cite{Shimakawa2017} when comparing lower and higher density regions of a protocluster at z$\sim$2.5. However, \cite{Shimakawa2017} found boosted SFRs for a given stellar mass among galaxies in the densest subset of their sample. By contrast, \cite{Hatch2011} found lower specific SFRs among members of two protoclusters at z$\sim$2.2 compared to the field. Finally, \cite{Koyama2013} saw redder colors among protocluster galaxies, whereas \cite{Hatch2011} found no such distinction between their protocluster and field groups.

The tendency for high redshift protocluster environments to host a larger fraction of high mass, actively star-forming galaxies differs from the trend seen in the local universe, and may indicate that star formation begins earlier in galaxies in rich environments \citep[e.g.,][]{Hatch2011, Ito2020}, consistent with theoretical predictions \citep[e.g.][]{Chiang2017}. Additional high redshift studies with complementary approaches to galaxy sample selection are needed to disentangle the influences that intrinsic and environmental differences have on the assembly histories of galaxies and clusters. Unique among the studies already mentioned, which select protocluster galaxies based on strong line emission or UV continuum emission, this work uses proximity to \lya\ nebulae. \lya\ nebulae (also known as Lyman-alpha blobs, LABs, or simply ``blobs'') are extended sources of \lya\ emission found at z$\sim$2-6. Typically more than 100 kiloparsecs across, these glowing clouds of gas are some of the largest objects known and have been associated with dense protocluster environments \citep[e.g.][]{Steidel2000, Matsuda2004, Saito2006, Prescott2008, Yang2009, Yang2010, Hennawi2015, Cai2017a, Cai2017b, Battaia2018}. Due to the expansion of the Universe, the ultraviolet light from \lya\ nebulae is redshifted to optical wavelengths, making it possible to observe these giant high-redshift structures with sensitive optical ground- and space-based telescopes. 

In order to probe the origin of environment-dependent trends in galaxy properties, we use these \lya\ nebulae as signposts of overdense regions and leverage high resolution Hubble Space Telescope (HST) observations of the GOODS-S extragalactic field to compare a composite group of galaxies in the vicinity of 6 known \lya\ nebulae to a control group in the field at z$\sim$2.3. We discuss data sources and quality requirements in Section~\ref{sec:data}, sample selection in Section~\ref{sec:samples}, and analysis methods in Section~\ref{sec:analysis}. We report our results and describe the various tests of the robustness of these findings in Section~\ref{sec:results}. We discuss our results and conclude in Sections~\ref{sec:discussion} and \ref{sec:conclusions}. Throughout, we assume the standard $\Lambda$CDM cosmology, i.e., $\Omega_m=0.27$, $\Omega_{\Lambda}=0.73$, $h=0.7$. The angular scale at $z=2.3$ is 8.4 kiloparsecs/arcsec (kpc/$^{\prime\prime}$). All magnitudes are in the AB system \citep{Oke1974}.

\section{Data}
\label{sec:data}

3D-HST is a Hubble Space Telescope (HST) spectroscopic survey covering four well-studied extragalactic fields: AEGIS, COSMOS, GOODS-S, and UDS \citep{Skelton2014}. It builds upon the Cosmic Assembly Near-infrared Deep Extragalactic  Legacy Survey (CANDELS), which obtained Hubble Wide Field Camera 3 (WFC3) imaging of the same regions, as well as the GOODS-N field \citep{Grogin2011}. The online 3D-HST catalog\footnote{3D-HST website and online catalog: https://3dhst.research.yale.edu/Data.php} provides imaging, photometry, spectra, emission line measurements, redshifts, and other derived properties for $\sim$100,000 galaxies in all five CANDELS fields using data from 3D-HST, CANDELS, and other space- and ground-based surveys \citep{Momcheva2016,Skelton2014,Brammer2012}. 

Building on CANDELS data, \cite{vanderwel2012} derived structural parameters for the galaxies in 3D-HST using the best S\'ersic model fits to CANDELS imaging. Their catalogs, available online\footnote{CANDELS structural catalogs: https://www2.mpia-hd.mpg.de/homes/vdwel/3dhstcandels.html}, complement those of 3D-HST and are line-matched using the same photometric IDs. \cite{Guo2013} provide additional data on objects in GOODS-S in their online catalog\footnote{CANDELS GOODS-S multi-wavelength catalog: https://cdsarc.cds.unistra.fr/viz-bin/cat/J/ApJS/207/24}, including limiting F160W magnitudes.

\subsection{Galaxy Properties and Quality Cuts} \label{sec:qualitycuts}

For all galaxies in our sample, we gathered WFC3 F160W $(H)$ and F814W $(I)$ fluxes from the 3D-HST GOODS-S photometric catalog (release V4.1), converting to AB magnitudes. We consider only those galaxies with good photometry \citep[i.e. only those whose $use\_phot$ flag was set to 1;][]{Skelton2014}. This is especially important given that over 90\% of sources near z=2.3 have only a photometric redshift, upon which we rely when selecting nebula-associated galaxies and their counterparts in the field. We also obtained the following stellar population parameters, which were estimated from the best fits to multiband photometry using FAST \citep{Kriek2009}: stellar mass, SFR, specific SFR, stellar age (since the onset of star formation), star formation decay timescale $\tau$, and dust attenuation A$_V$. To ensure we were only using galaxies with high quality SED fits, we required the reported $\chi^2$ of the FAST fit to be less than 2. This cut retained 79\% of the GOODS-S photometric galaxy sample.

From the 3D-HST GOODS-S grism catalog (release V4.1.5), we collected fluxes and equivalent widths for the OII, OIII, and H$\beta$ emission lines, requiring S/N$>$2. The ``z\_best'' grism catalog provides redshifts as well as combined UV+IR star formation rates. The ``best'' redshift available may be a spectroscopic redshift or, in its absence, a grism redshift. In the absence of both, a photometric redshift is provided \citep{Momcheva2016}. UV+IR star formation rates based on fluxes with low S/N were rejected by requiring the SFR flag to be set to zero. 

Finally, we obtained S\'ersic indices $n$ and effective radii r$_e$ from \cite{vanderwel2012}. We required the provided flag to be set to zero, indicating a good fit. We also cross-matched objects in the 3D-HST GOODS-S catalogs with that of \cite{Guo2013} to associate the F160W limiting magnitude of the surrounding region to each galaxy in our sample.

Given data on over 50,000 sources within the GOODS-S field, we proceeded to compare a range of properties between galaxies in the vicinity of \lya\ nebulae and similarly selected galaxies in the field. 

\section{Sample Selection}
\label{sec:samples}

\subsection{Ly$\alpha$ Nebula Sample}
\label{sec:nebulasample}

Our \lya\ nebula sample is taken from a blind survey of four extragalactic fields \citep{Yang2010}. The nebulae were selected using a narrowband filter corresponding to z=2.3${\pm}$0.037. Based on the derived number density from this blind search, the authors estimated a halo mass hosting a \lya\ nebula to be $\geq$10$^{13}$ M$\odot$. Of the 25 \lya\ nebulae detected, sixteen lie within the Chandra Deep Field South (CDFS), and six of these overlap the 3D-HST GOODS-S catalog.

\begin{figure}
     \centering
     \subfloat{%
        \includegraphics[width=0.52\textwidth]{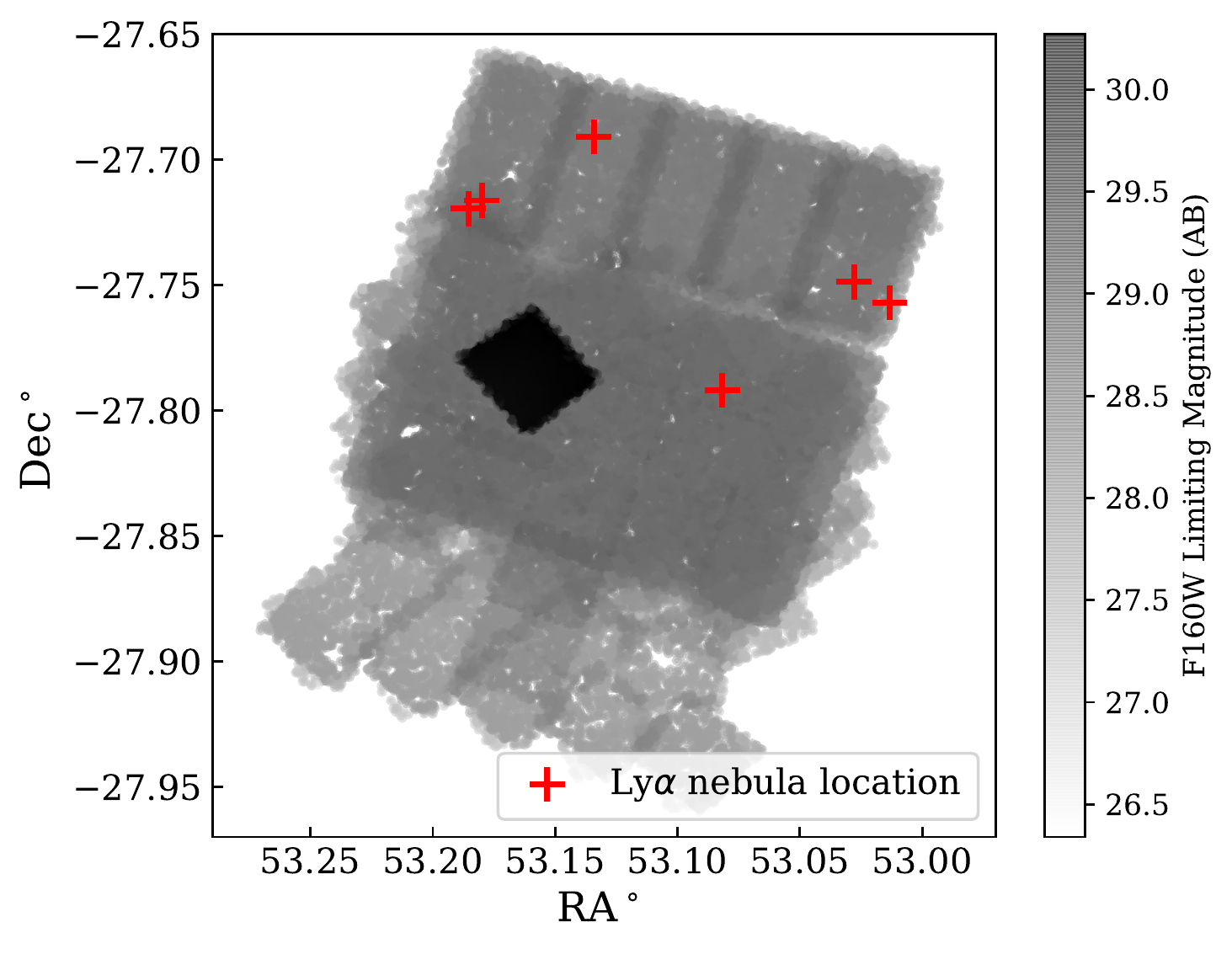}%
        \label{}%
        }\qquad
     \subfloat{%
        \includegraphics[width=0.37\textwidth]{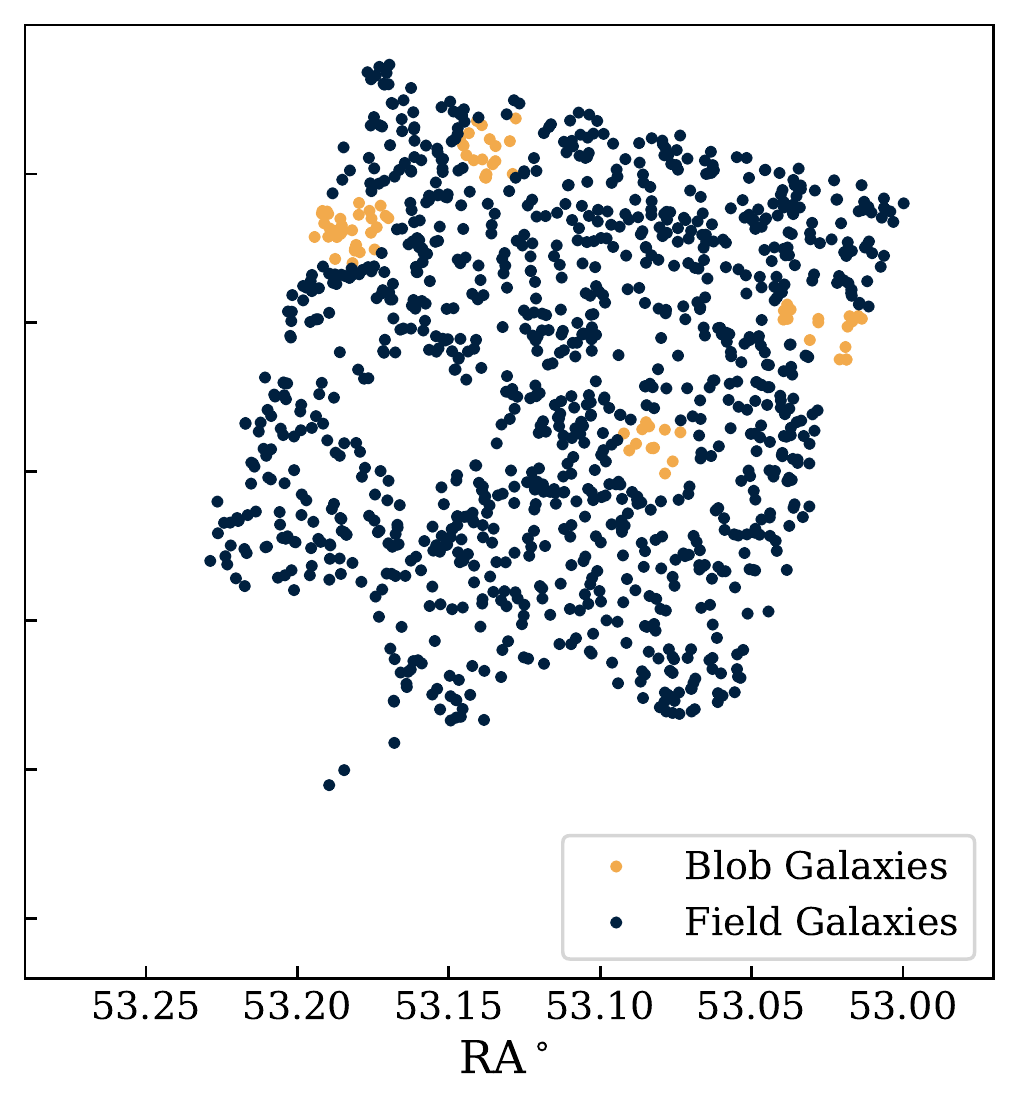}%
        \label{}%
        }
        \caption{{\emph{Left:} The GOODS-S extragalactic field ($\sim$170 arcmin${^2}$), showing F160W limiting magnitude of galaxies in shades of gray \citep{Guo2013}, and locations of \lya\ nebulae as red crosses \citep{Yang2010}. \emph{Right:} Selected blob ($yellow$) and field ($black$) galaxies at z$\sim$2.3. Galaxies in the shallower lower third of the field and in the deeper Hubble Ultra Deep Field  \citep{Koekemoer2013, Beckwith2006}} are eliminated after imposing a limiting magnitude constraint (Section \ref{sec:galsample}).}
        \label{fig:GOODSS_Hlim_both}
\end{figure}

\begin{figure}
\centering
\includegraphics[width=0.5\textwidth]{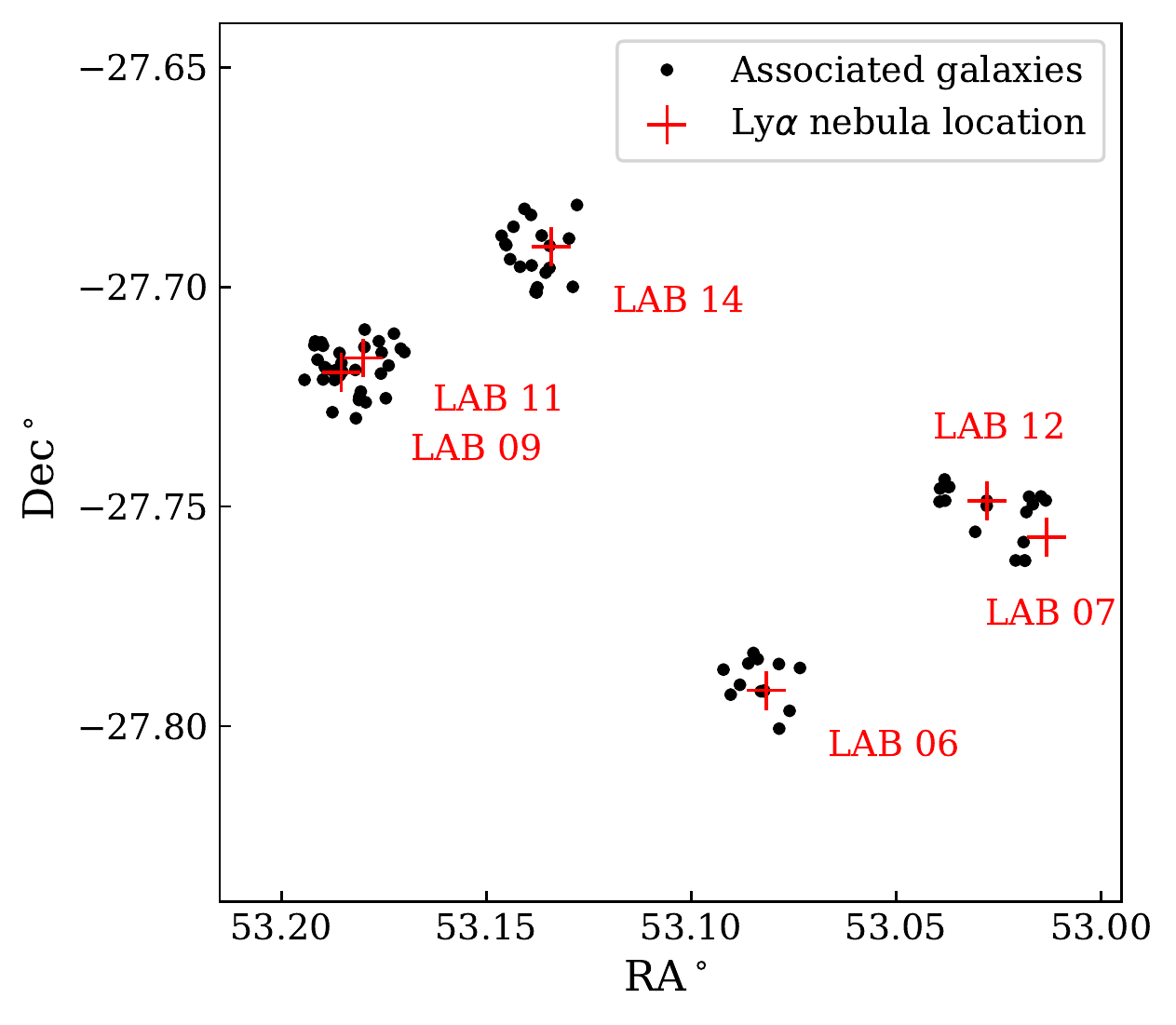}
\caption{{Close-up view of the blob galaxy sample. The blob locations are labeled with red crosses, and associated galaxies (as defined in Section \ref{sec:galsample}) are indicated as black circles.} \label{fig:GOODSS_BlobGals}}
\end{figure}

\subsection{Galaxy Sample} \label{sec:galsample}

We defined \lya\ nebula-associated galaxies to be those within 40 arcseconds of any of the six \lya\ nebulae locations and with z=2.3$\pm$0.15, forming a composite ``blob galaxy'' sample. We selected these radius and redshift ranges based on a number of considerations. The redshift range within which we can expect to find galaxies associated with a nebula must be large enough to capture true members, given the redshift uncertainties, but not so large as to include a great number of interlopers. The best available redshift measurements for the vast majority of galaxies near z=2.3 in the 3D-HST catalog are photometric, with typical uncertainties of $\Delta$z=0.15, corresponding to $\Delta$z/(1+z)=0.05 at this redshift, roughly an order of magnitude larger than any $\Delta$z expected from peculiar motion (even considering cluster members). 

A similar balance must be struck when defining a radius around each Ly$\alpha$ nebula within which we select blob galaxies. Thanks to space-based imaging, uncertainty in on-sky position is negligible. From our analysis, a 40" radius ($\sim$320 proper kpc at z$\sim$2.3) emerged as the distance beyond which statistically significant differences between blob and field galaxies disappear (see Section \ref{sec:centralblobgals}). Finally, survey depth varies in GOODS-S (Figure \ref{fig:GOODSS_Hlim_both}, \emph{left}), so in order to make a fair comparison of the galaxy population across the field, we rejected both shallower and deeper regions by only selecting galaxies with an associated F160W limiting magnitude between 28.25 and 28.75 (Figure \ref{fig:GOODSS_Hlim_both}, \emph{right}). However, we note that when we do not correct for field depth in this way, we see qualitatively the same results. 

Eighty-six galaxies make up the composite blob sample (Figure \ref{fig:GOODSS_BlobGals}). We defined our control sample as all other GOODS-S sources lying outside of the defined blob regions whose redshifts and limiting magnitudes fell within the same specified ranges, yielding a population of 1,099 galaxies. In total, 1\% of the galaxy redshifts used in this work are are spectroscopic, 7\% are grism, and 91\% are photometric. Only a fraction of the blob and field galaxies have data for each property of interest that meet the quality requirements outlined in Section \ref{sec:qualitycuts}. Thus, the number of galaxies $n$ included in a given analysis is specified where applicable.

\section{Analysis} 
\label{sec:analysis}

We utilized two non-parametric statistical tests in order to compare properties between blob and field galaxies: the 2-sample Kolmogorov–Smirnov (KS) test and the k-sample Anderson-Darling (AD) test, implemented using the $ks\_2samp$ and $anderson\_ksamp$ functions in SciPy's $stats$ module \citep{Virtanen2020}. Given two sets of observations of a single, continuous variable, these tests compare the distributions of the two samples without assuming anything about the nature of the parent distribution. Graphically, both methods arrange the data from each sample into empirical cumulative probability distributions (CPDs). The KS test reports the largest difference between the two CPDs at any one interval, and from this statistic a $p$-value between 0 and 1 is calculated given the size $n$ of each sample \citep{Hodges1958}. Meanwhile, the AD test calculates the sum of the squared differences between the CPDs at \emph{every} interval, and gives more weight to differences in the tails of the distributions. As with the KS test, a $p$-value is produced based on this statistic and the sample size \citep{Scholz1987}. In both cases, the reported $p$-value indicates the probability that such a difference would be seen by chance assuming that the two samples are drawn randomly from the same parent distribution (the null hypothesis). A low $p$-value, resulting from a large difference between the CPDs, indicates that we have reason to reject the null hypothesis. For example, a $p$-value of $<$0.05  indicates that there is less than a 5\% chance that two given samples come from the same population. This is a commonly accepted threshold for statistical significance that is used in works comparable to this one \citep[e.g.][]{Hatch2011, Cooke2014, Shimakawa2017}, and is what we adopt here.

\section{Results} 
\label{sec:results}

\subsection{Fiducial Results}

\begin{figure}
\includegraphics[width=0.69\textwidth]{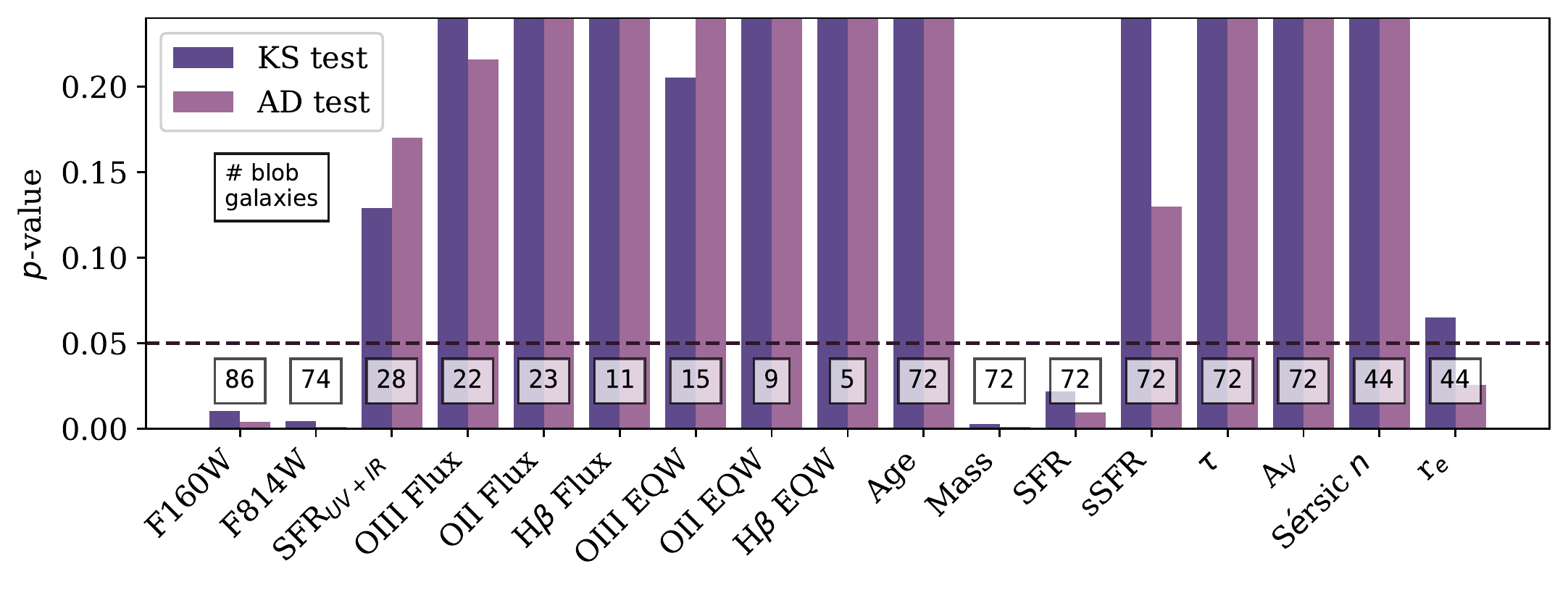}
\centering
\caption{{KS (\emph{dark purple}) and AD (\emph{light purple}) test $p$-values for all properties compared between galaxies near \lya\ nebulae and those elsewhere in the GOODS-S field at z$\sim$2.3 (see Section \ref{sec:qualitycuts}). Both tests indicate statistically significant differences between the two groups in magnitude (F160W and F814W), stellar mass, star formation rate (SFR), and effective radius (r$_e$). Among the other traits, $p$-values from both tests are well above the $p$=0.05 threshold (\emph{dashed line}).} \label{fig:allpvals_radius40}}
\end{figure}

\begin{figure*}
\includegraphics[width=0.69\textwidth]{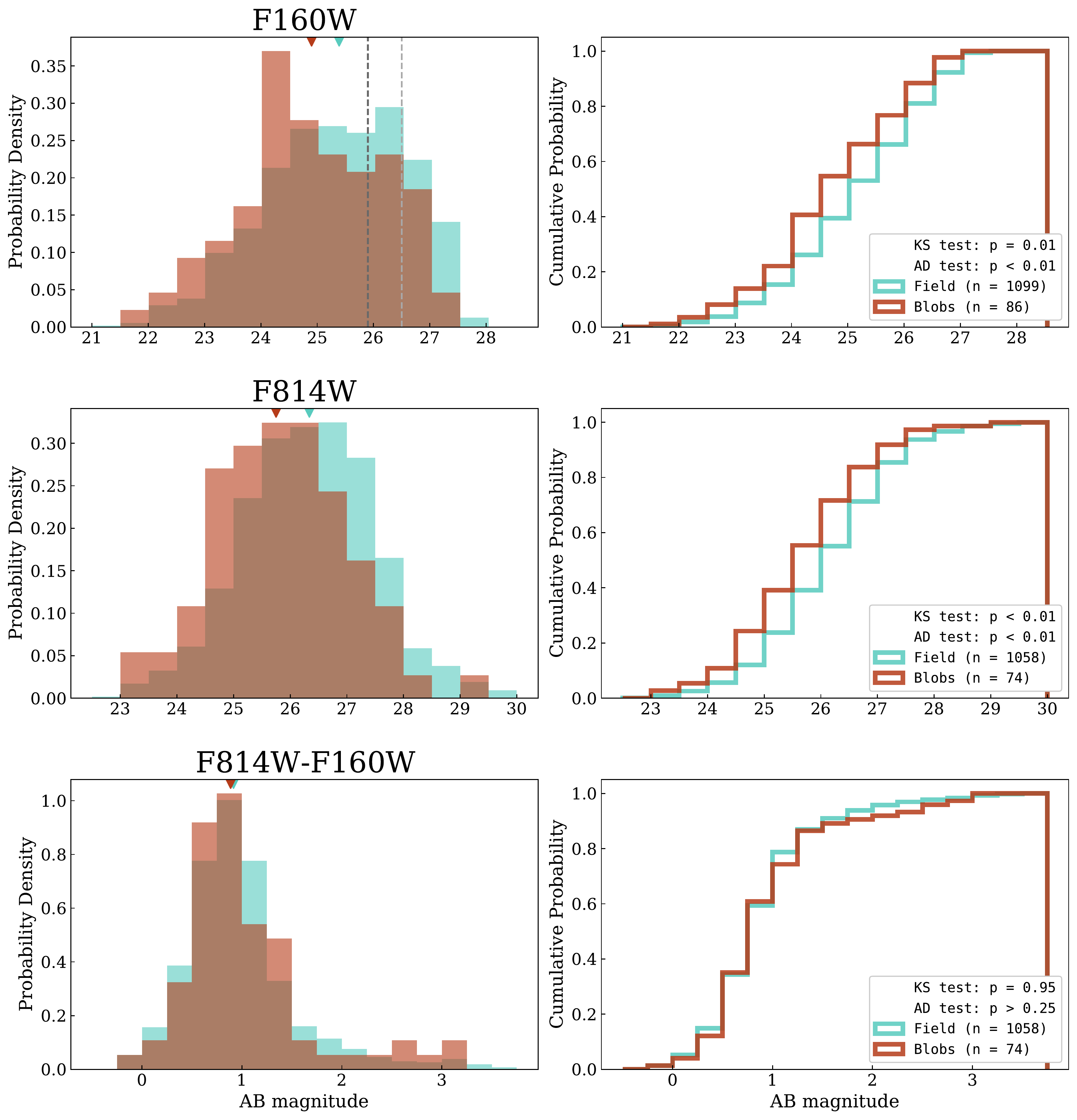}
\centering
\caption{{Comparisons of the distributions of galaxies near \lya\ nebulae (\emph{red}) and those in the field (\emph{blue}) for three properties: F160W mag, F814W mag, and F814W-F160W color index. The 75\% and 50\% F160W completeness thresholds are shown as dark and light gray dashed lines, respectively. The median value of each group is indicated with a triangle. Sample sizes and $p$-values from the KS and AD statistical tests are provided in the inset boxes. In both filters, a systematic offset towards brighter magnitudes is seen among the blob galaxies with respect to those in the field, with correspondingly low $p$-values. No significant difference is seen in F814W-F160W color.} \label{fig:hists_filters}}
\end{figure*}

\begin{figure*}
\includegraphics[width=0.68\textwidth]{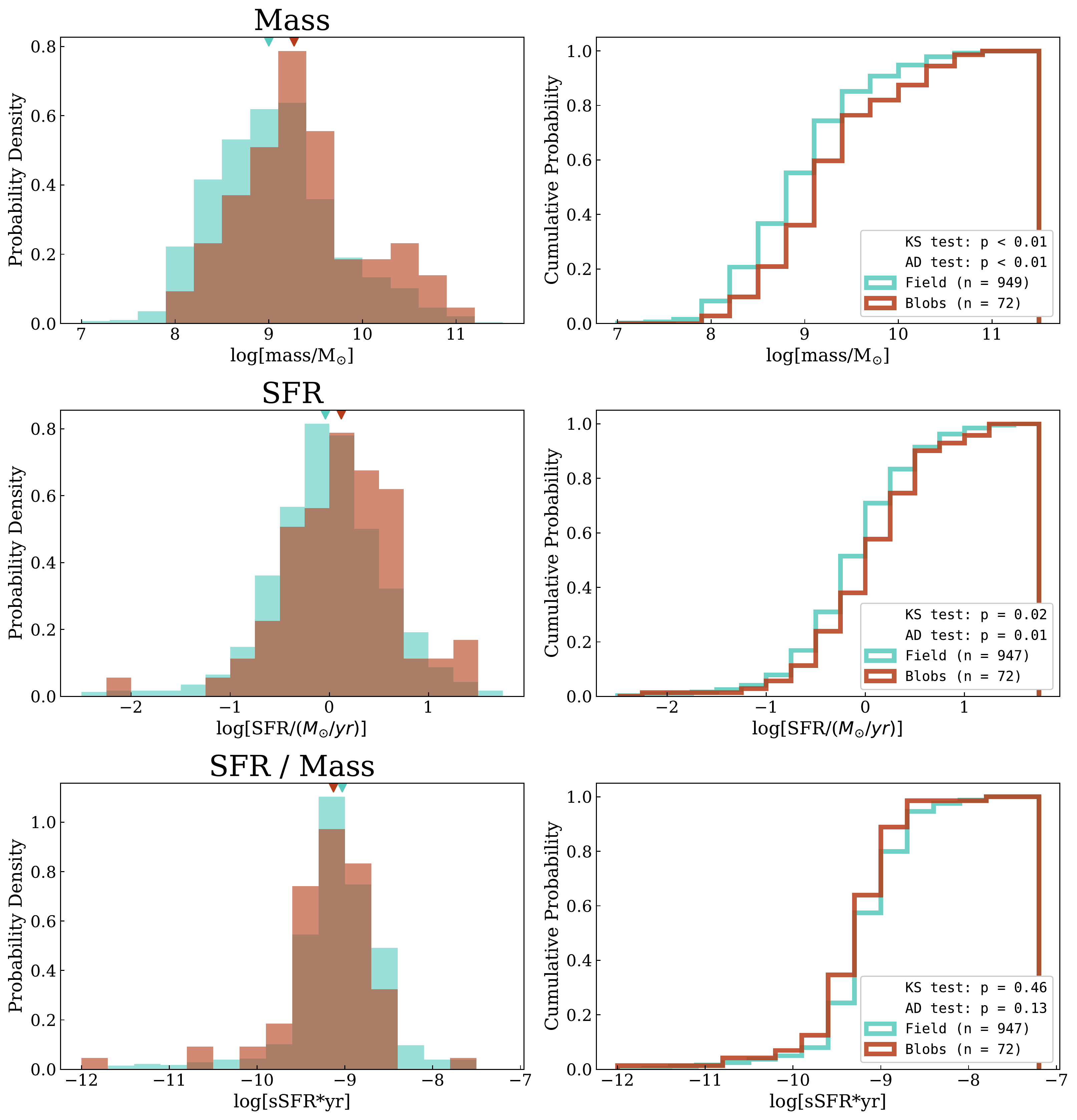}
\centering
\caption{{Comparisons of the distributions of galaxies near \lya\ nebulae (\emph{red}) and those in the field (\emph{blue}) for three properties: stellar mass, star formation rate (SFR), and SFR per unit stellar mass (or specific star formation rate, sSFR). The median value of each group is indicated with a triangle. Sample sizes and $p$-values from the KS and AD statistical tests are provided in the inset boxes. A systematic offset towards higher mass and SFR is seen among the blob galaxies with respect to those in the field, with correspondingly low $p$-values. No statistically significant difference is seen in sSFR between the groups.} \label{fig:hists_FAST}}
\end{figure*}

\begin{figure*}
\includegraphics[width=0.68\textwidth]{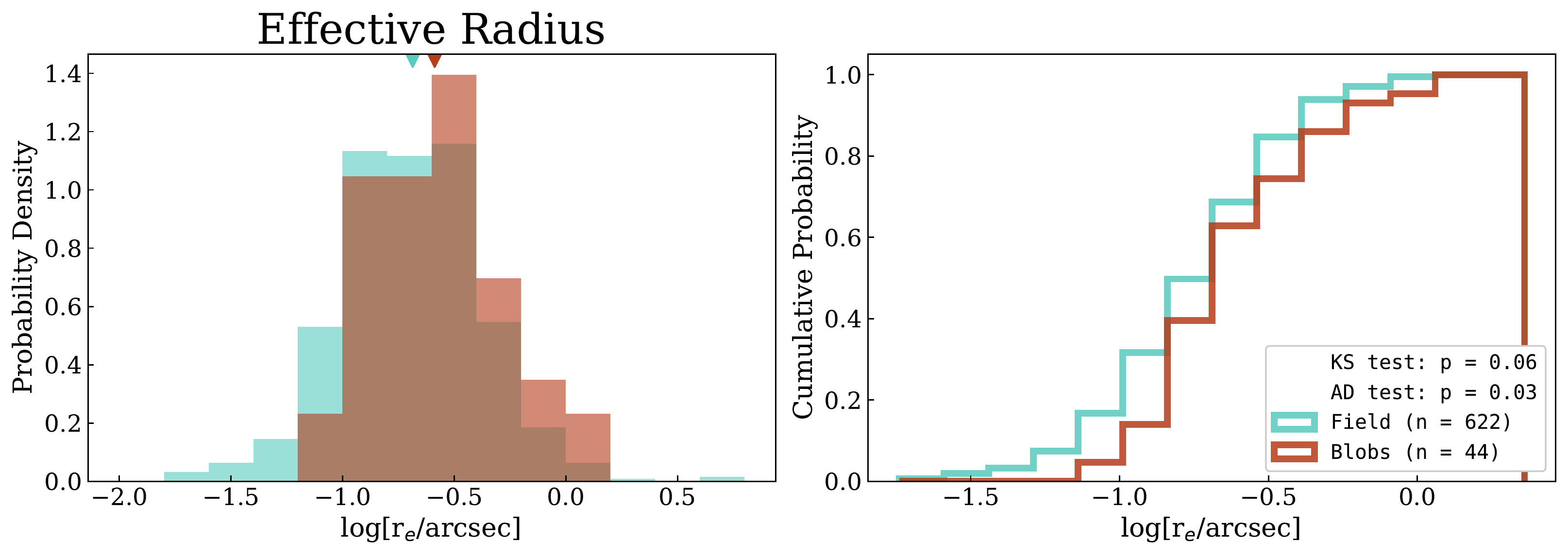}
\centering
\caption{{Comparison of blob (\emph{red}) and field (\emph{blue}) galaxy effective radius distributions. The median value of each group is indicated with a triangle. Sample sizes and $p$-values from the KS and AD statistical tests are provided in the inset boxes. A systematic offset towards larger effective radii is seen among the blob galaxies with respect to those in the field, accompanied by correspondingly low $p$-values.} \label{fig:hists_re}}
\end{figure*}

Resulting $p$-values from both statistical tests are displayed for every property we examined in Figure \ref{fig:allpvals_radius40}, with the differential and cumulative distributions of selected properties shown in Figures \ref{fig:hists_filters}, \ref{fig:hists_FAST}, and \ref{fig:hists_re}. We see statistically significant differences between the blob and field galaxies in five properties: F160W and F814W magnitudes, stellar masses, star formation rates (SFR), and effective radii (r$_e$). In these five cases, the size of the blob galaxy sample is around 75, and the size of the field galaxy sample is around 1000, except for r$_e$, which has sample sizes $\sim$30\% smaller. The differences seen in this group of properties are well distinguished from the other galaxy trait comparisons we performed: in all five cases where the $p$-value from the KS test falls below the $p$=0.05 threshold, we see that the $p$-value from the AD test does as well (we include r$_e$ in this group, with KS test $p$=0.06, AD test $p$=0.03). In all other cases, the $p$-values from both tests fall well above this threshold. It is important to note sample sizes in each case, as these vary widely based on the availability and quality of data. Properties which have a blob galaxy sample smaller than $\sim$30, such as the emission line measurements, have too few blob galaxies for a reliable KS test result \citep{Razali2011}, meaning that if a true difference were present, we would be unable to detect it with so few galaxies. As a concrete example, unlike the FAST-derived SFR measurement where the blob galaxy sample size is $n$=72 (KS test $p$=0.02), we have only 28 quality blob galaxy measurements for the combined UV+IR SFR (SFR$_{UV+IR}$), with KS test $p$=0.13. When we compared the FAST SFR of these same blob galaxies ($n$=28) to the field, the resulting KS test shows no significant difference ($p$=0.19), even though we know that with a larger blob galaxy sample size ($n$=72), a significant difference is indeed detected ($p$=0.02) (Figure \ref{fig:allpvals_radius40}).

Our most robust result is that blob galaxies are brighter in both filters than those in the field. Normalized histograms comparing the filter magnitudes are shown in Figure \ref{fig:hists_filters}, with blob galaxy distributions in $red$, field galaxy distributions in $blue$, and sample sizes and KS/AD test $p$-values shown in the inset boxes. In both filters, we see a proportionally larger number of bright blob galaxies and a systematic offset towards brighter magnitudes in the blob galaxy group with respect to the field. It must be noted that the distributions are incomplete at the faint end; the 50\% and 75\% F160W completeness thresholds for GOODS-S are indicated with light and dark gray dashed lines respectively. However, given that both samples are depth-corrected, the two groups should be incomplete in the same way, so, all else being equal, we would expect the distributions to fall off at the faint end in the same manner. Instead, they do not: in F160W the offset to brighter magnitudes becomes apparent before 26.5 mag, the 50\% completeness threshold for this filter in this region. The same type of offset is seen also in the F814W data. It does appear, then, that blob distributions have proportionally more bright galaxies, as well as fewer faint ones. At the same time, while blob galaxies appear to be systematically brighter than those in the field, they do not exhibit a statistically significant difference in F814W-F160W color (Figure \ref{fig:hists_filters}).

Unsurprisingly, given their systematically brighter magnitudes, blob galaxies appear to be consistently shifted towards higher stellar masses and SFRs relative to galaxies in the field when we examine these FAST-derived properties (Figure \ref{fig:hists_FAST}). No such offset is seen when comparing SFR per unit stellar mass (specific SFR) between the two groups, suggesting that while the blob group is skewed towards higher masses, the two groups may follow the same SFR-mass relation; we return to this point in Section \ref{sec:discussion}.

Finally, the effective radii of blob galaxies appear to be systematically larger than those in the field. The corresponding histograms are shown in Figure \ref{fig:hists_re}.

Thus far, we have demonstrated that there is a statistical difference between blob and field galaxies in these five key properties. Next, we aim to quantify the nature and magnitude of these differences. The non-parametric KS and AD tests are sensitive to differences in shape, center and spread between two distributions, but the resulting statistic and $p$-value do not indicate which one (or more) of these differences is at play, or how large it is. If our visual impression that the blob distributions are systematically offset from the field is correct, with the primary difference being their centers and not their shape or spread, then we can directly compare their medians. We compare medians instead of means to avoid being sensitive to a few outliers in our SFR and r$_e$ data. For a given property we added the difference between the blob and field galaxy medians to every individual blob galaxy measurement, thereby preserving the blob distribution's shape and spread but shifting its median to coincide with that of the field. When we redo the KS and AD tests, we find that for all five properties (F160W, F814W, stellar mass, SFR, and effective radius), the KS test $p$-values rise to 0.96, 0.94, 0.70, 0.67, and 0.72 respectively, and the AD test $p$-values exceed 0.25. This indicates that the shape and spread of the distributions do not differ significantly, and that the difference in median is driving the statistical difference. 

Thus, this simple exercise indicates that galaxies near blobs are roughly 0.49 magnitudes or 1.6 times brighter in F160W and 0.59 magnitudes or 1.7 times brighter in F814W, have 1.9 times more stellar mass, have star formation rates that are 1.4 times higher, and have effective radii that are 1.3 times larger than galaxies in the field. In what follows, we explore whether these results could be influenced by issues of incompleteness or sample selection.

\subsection{Verifying the Results}
\label{sec:verifying}

In this section we describe the various measures we took to test the reliability of our results. In particular, we investigated the consequences of imposing completeness limits, how the size of the region within which we select blob galaxies affects our results, and whether a subset of the six blob regions is primarily responsible for the outcome. We also estimated the expected false positive rate by selecting six random blob-free locations within GOODS-S and analyzing the resulting ``mock'' blob and field galaxies in identical fashion. 

\subsubsection{Sample Completeness}

The faint end of our galaxy sample will necessarily be affected by incompleteness. However, given that our sample is drawn from regions of consistent survey depth (F160W limiting magnitude m$_{lim}$ = 28.50${\pm}$0.25 mag), we expect our blob and field galaxy populations to be incomplete to the same degree, allowing them to be fairly compared with one another while keeping in mind this limitation. Nevertheless, we confirmed that our qualitative results are robust to completeness issues via two distinct experiments: (1) removing the faintest galaxies and (2) correcting the sample for incompleteness by adding simulated galaxies to the faintest bins. In what follows, we use the completeness fraction as a function of F160W magnitude for the GOODS-S field \citep{Skelton2014}.

\begin{figure}
\includegraphics[width=0.6\textwidth]{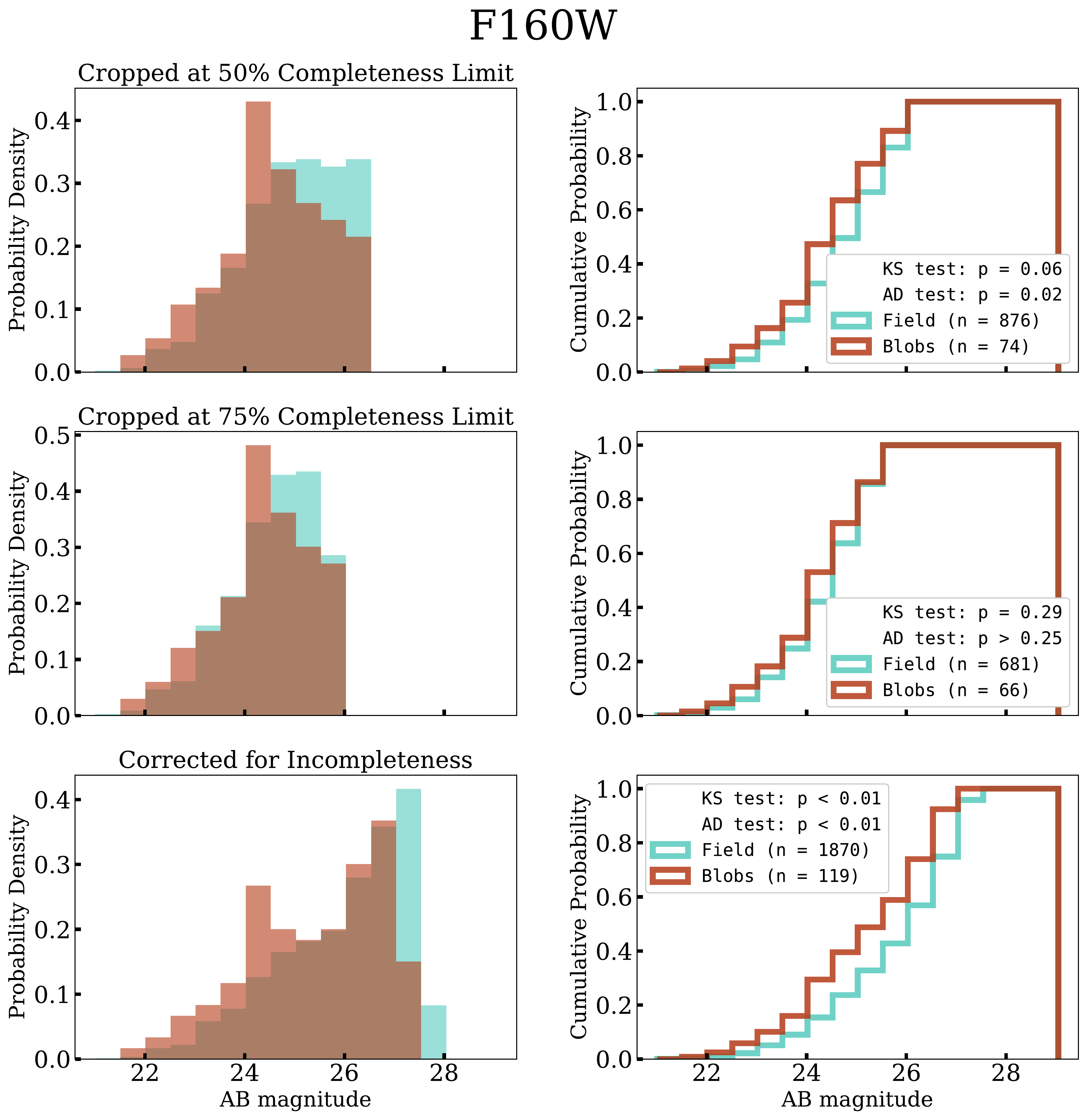}
\centering
\caption{{Comparison of blob and field galaxy F160W distributions when cropping the sample at the 50\% (\emph{top}) and 75\% (\emph{middle}) completeness limits. Alternatively, we simulated a sample corrected for incompleteness by adding the appropriate number of mock galaxies to the faintest bins (\emph{bottom}).} \label{fig:completeness}}
\end{figure}

First, when only considering blob and field galaxies with F160W magnitudes brighter than the 50\% completeness limit, the AD and KS $p$-values remain significant at 0.02 and 0.06 respectively. When we crop the sample at the 75\% completeness limit, both $p$-values rise above 0.25 (Figure \ref{fig:completeness}, \emph{top} and \emph{middle} panels). As discussed in Section \ref{sec:results}, though our sample is corrected for depth, the blob and field galaxy F160W and F814W distributions do not fall off in the same manner at the faint end as would be expected, but rather there appear to be proportionally fewer faint galaxies in the blob regions (Figure \ref{fig:hists_filters}). When we remove the faintest galaxies to increase the completeness of our sample, the signal weakens, emphasizing that the faint end is contributing to the differences we detect.

In the second approach, we simulated a complete F160W sample. Using \citeauthor{Skelton2014}'s (\citeyear{Skelton2014}) completeness fraction for each half-magnitude bin, we calculated the number of missing galaxies in a given bin, and drew a corresponding number of galaxy magnitudes randomly, assuming a uniform distribution within the bin. After artificially recovering the faintest galaxies in our sample in this way, differences between the blob and field F160W distributions become even more pronounced, with both AD and KS $p$-values falling below 0.01 (Figure \ref{fig:completeness}, \emph{bottom} panel).

These incompleteness tests suggest that the existing data are already revealing real differences between blob and field galaxies. In the remainder of the paper, we therefore choose to stay close to the data, focusing on the fiducial depth-corrected sample. We leave a more in-depth investigation of the impact of incompleteness on all other measured quantities to future work.

\subsubsection{The Blob Region Radius}
\label{sec:blobregionradius}

\begin{figure}
\includegraphics[width=0.69\textwidth]{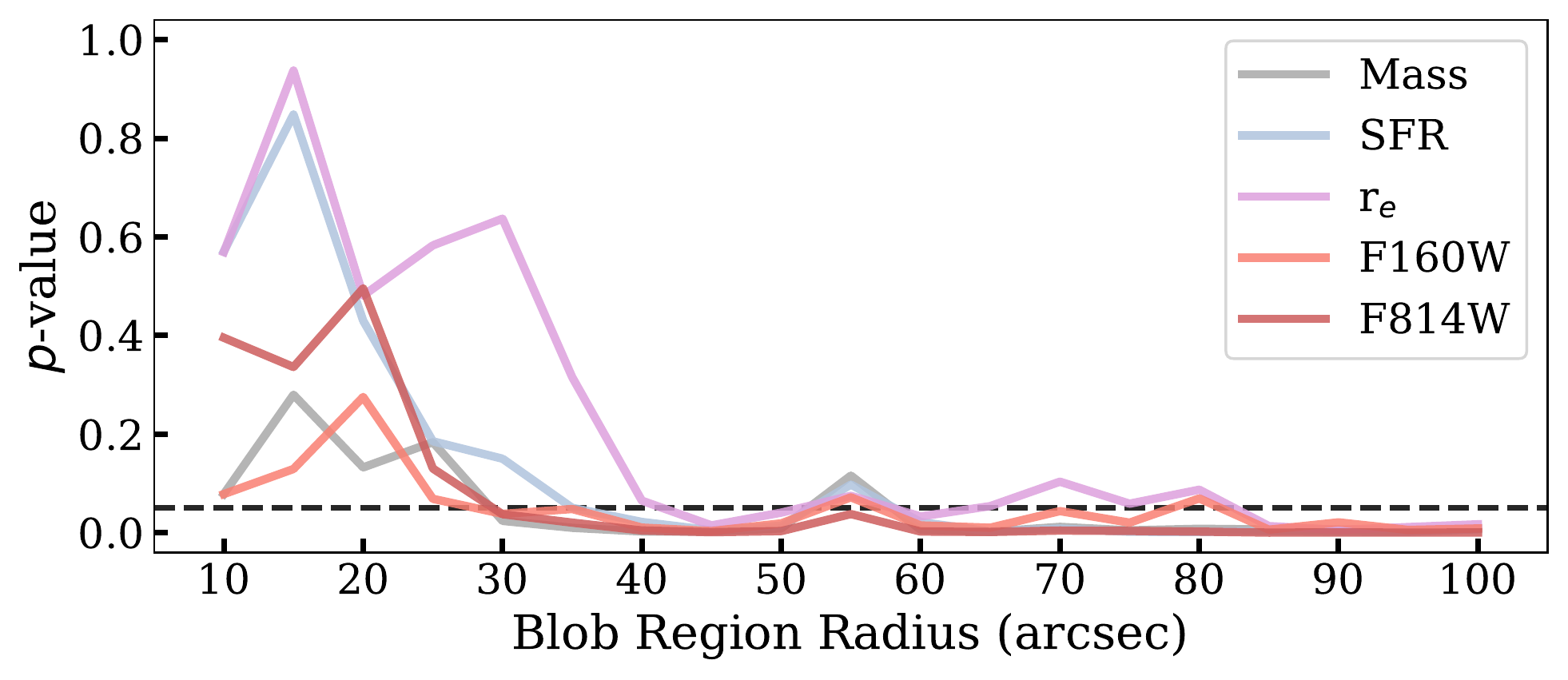}
\centering
\caption{{KS test results comparing selected properties of sets of blob and field galaxies for increasing blob region radius. It is not until the defined blob region reaches $\sim$40" in radius that the blob-associated galaxy sample size is large enough for the KS test to detect the difference between galaxies in the blob region and those in the field.} \label{fig:GOODSS_pvalcurve}}
\end{figure}

In using \lya\ nebulae as markers of overdensities, we must define the angular extent of such regions. A small ``blob region'' radius, centered on a \lya\ location, is more likely to capture a high fraction of true inhabitants of the dense environment, but restricts our sample size. As the radius is extended, we benefit from a larger blob galaxy sample, but that sample is more likely to contain a greater fraction of interlopers - galaxies that are not subject to the same potential influences as those embedded in an overdense environment. Making no \emph{a priori} assumptions about what the plausible physical size of such a dense region might be, we tested blob region radii from 10"-100" ($\sim$84-840 kpc). This range straddles the expected virial radius of $R_{vir}$=219 kpc for a galaxy group mass halo ($\sim$10$^{13}$ M$\odot$) at z$\sim2.3$, based on a standard spherical collapse calculation \citep{Bryan1998}. In steps of 5", we compared the properties of each new set of trial blob and field galaxies using the KS and AD tests as before. This allowed us to see how the $p$-values changed as the defined blob region varied in size. For simplicity and as the more conservative option we plot only the KS test $p$-values in this subsection (Figures \ref{fig:GOODSS_pvalcurve}-\ref{fig:GOODSS_pvalcurves_100randomlocs}). However, in general we found the AD test to produce lower $p$-values given the same samples.

In Figure \ref{fig:GOODSS_pvalcurve} we see that comparisons in the five selected properties produce $p$-values that dip below $p$=0.05 when the blob region reaches $\sim$40" in radius. For r=10-35", the blob galaxy sample size grows from $\sim$12 to $\sim$65 when considering the filter magnitudes; from 11 to 61 when considering the mass and SFR; and from 8 to 35 when considering the effective radius. The larger $p$-values seen in this range show the limit of the KS test for small sample sizes - we have demonstrated that the two populations differ in these properties when r=40" (e.g. Figure \ref{fig:allpvals_radius40}), but these differences cannot be detected at smaller radii given the limited blob galaxy sample size.

A potential problem with varying the size of the blob region radius in this way is that as the blob galaxy sample grows the field galaxy sample shrinks, with members of the latter group switching to the former as the blob region radius increases. Consequentially, the blob sample is not being compared to a static group, but rather one whose distribution changes slightly at every step. In order to control for this, we performed the same test but this time defined field galaxies to be those that lie greater than 60" from any \lya\ nebula location, giving us a consistent field group to which to compare blob galaxies. As we scanned through blob region radii of 10", 20", 30", 40", and 50", those galaxies that fell in the gap between the blob region radius and our field galaxy radius were simply ignored. This method produced results that were virtually identical to Figure \ref{fig:GOODSS_pvalcurve}, demonstrating that the impact of slight changes in the field galaxy distribution is negligible when using the previous approach.

\subsubsection{The Core Blob Galaxies}
\label{sec:centralblobgals}

\begin{figure}
    \centering
    \includegraphics[width=0.5\textwidth]{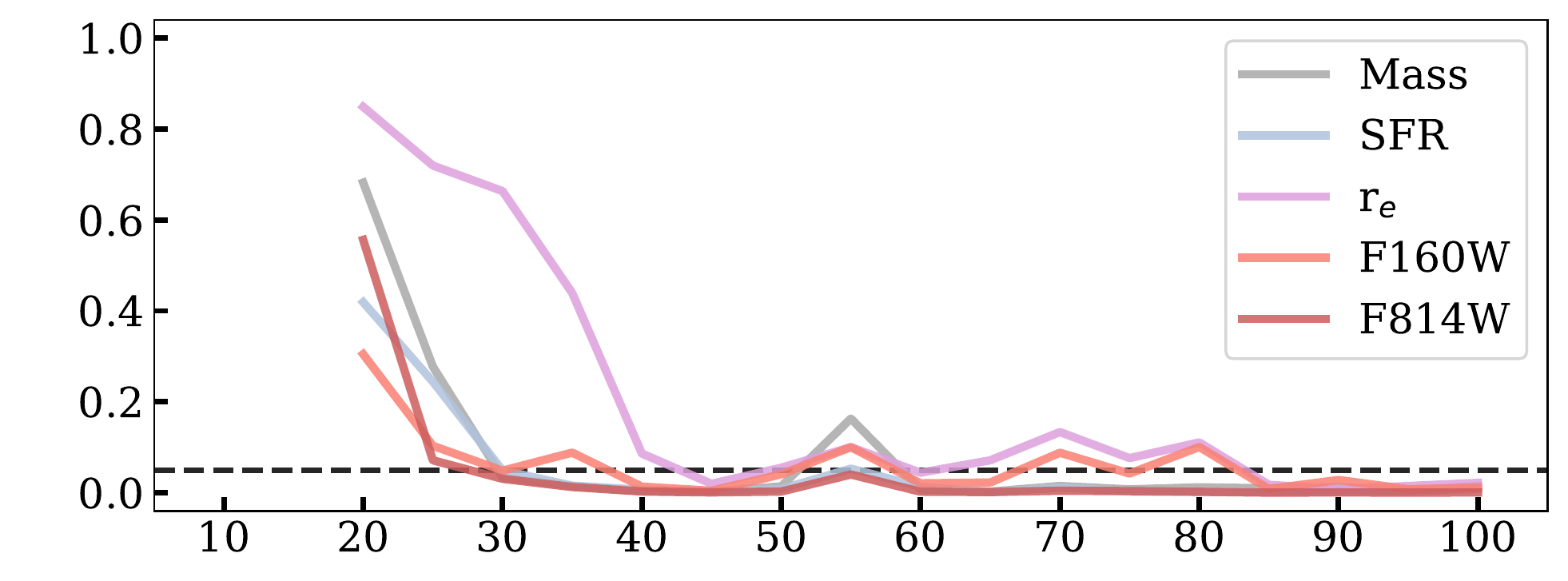}
    \includegraphics[width=0.5\textwidth]{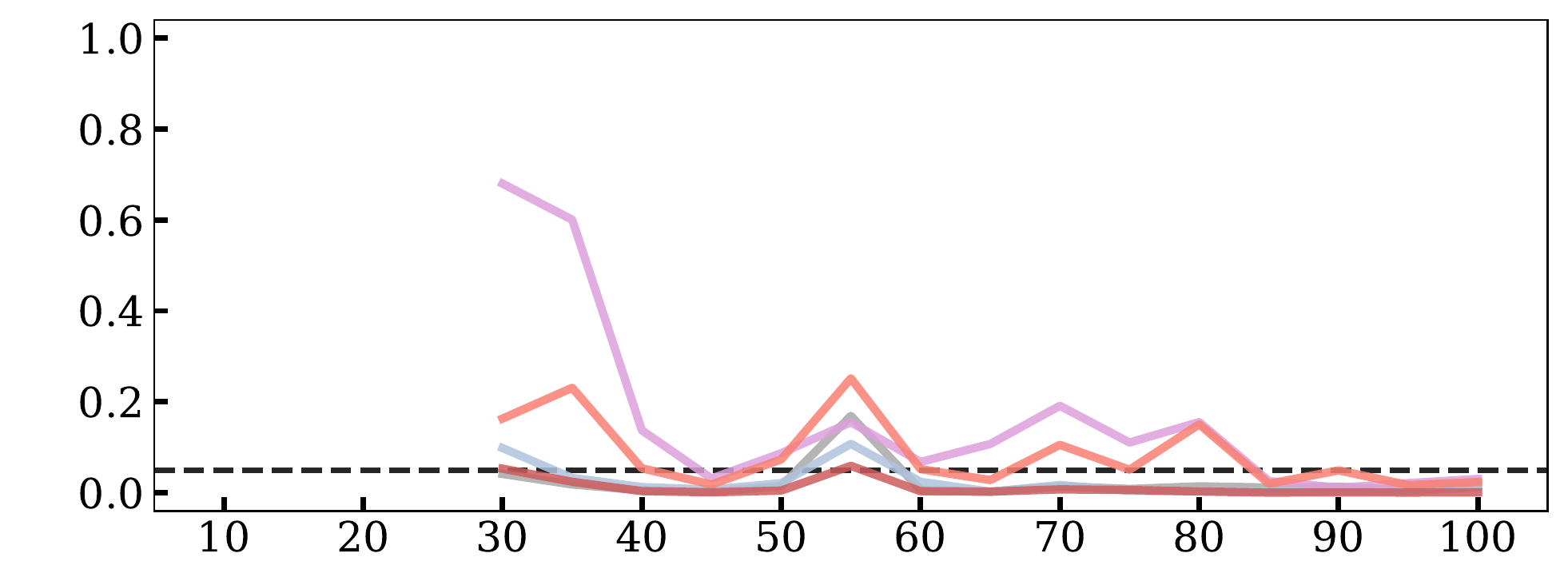}
    \includegraphics[width=0.5\textwidth]{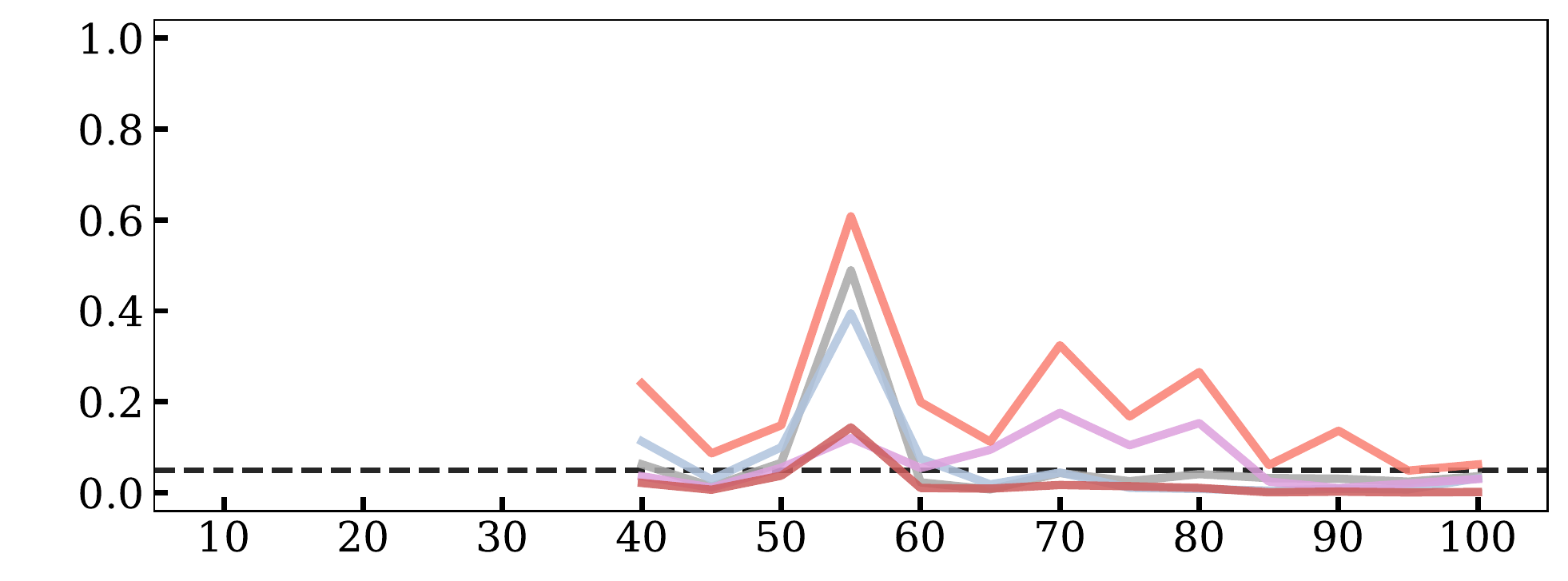}
    \includegraphics[width=0.5\textwidth]{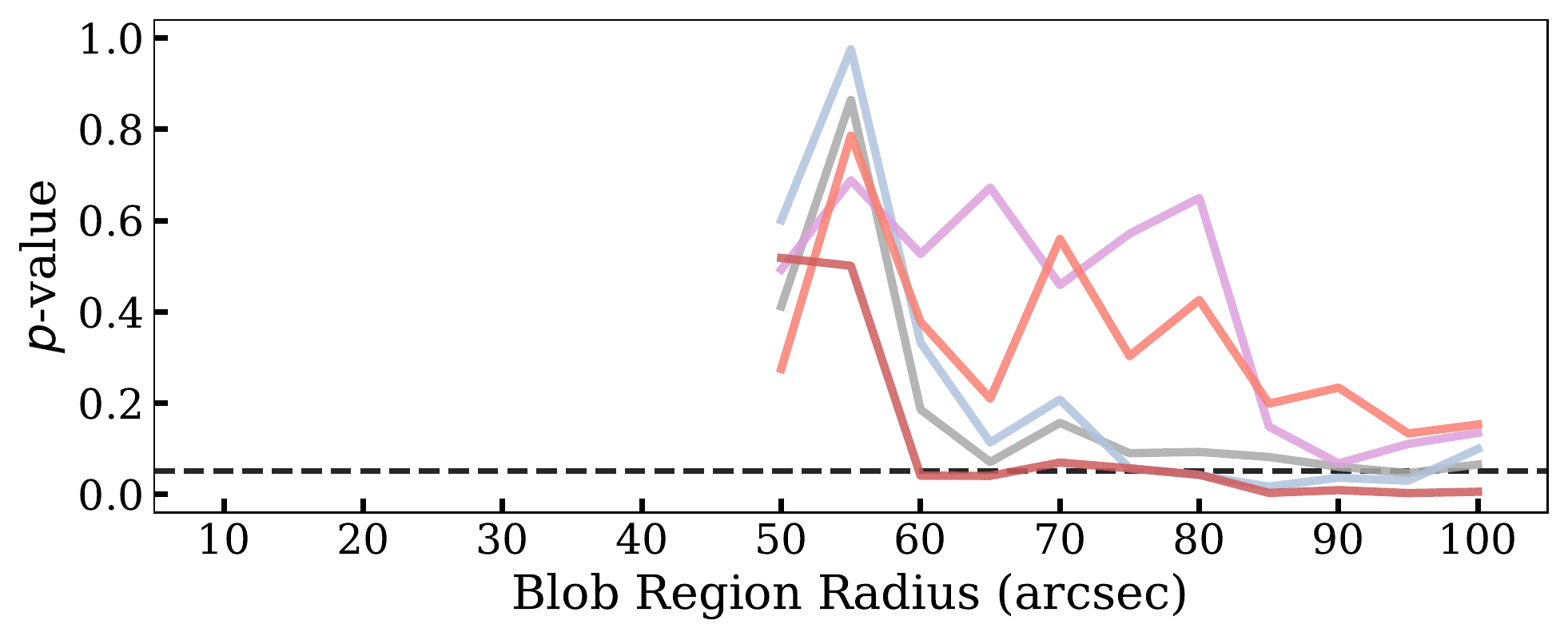}
    \caption{{Resulting $p$-value curves when galaxies within 10" (\emph{first panel}), 20" (\emph{second}), 30" (\emph{third}), and 40" (\emph{fourth}) of any blob are absent from the analysis. When galaxies within a 40" blob region radius are missing (\emph{fourth panel}), statistical differences in the five selected properties disappear, indicating that galaxies within this radius are driving the differences we detect, and not those at larger distances from the \lya\ nebulae.}}
    \label{fig:GOODSS_pvalcurves_nocentralgals}
\end{figure}

Given that the $p$-values for the five properties in Figure \ref{fig:GOODSS_pvalcurve} dip below and generally stay under $p$=0.05 from r=30" out to r=100", we then investigated whether 1) the galaxies closest to the blobs were responsible for this outcome, differing from the field enough to depress the $p$-values at large blob region radii despite the likely possibility that larger radii will incorporate more and more interlopers into the blob galaxy sample, or whether 2) galaxies at relatively large distances from the blobs were contributing to this effect. To this end, we progressively removed the most central blob galaxies from the analysis entirely, and, allowing the blob region radius to vary as before, compared the properties of the blob and field galaxy subsets. The results of this test are shown in Figure \ref{fig:GOODSS_pvalcurves_nocentralgals}. With more and more central blob galaxies missing, the $p$-value curves steadily rise, demonstrating that the core blob galaxies are indeed responsible for the low $p$-values seen in Figure \ref{fig:GOODSS_pvalcurve}. When central galaxies within 40" of any blob are absent, statistically significant differences in the five selected properties disappear. This is not due to prohibitively small sample sizes, as the blob sample size at this stage (those galaxies with 40"$<$r$<$50") is large enough (n=43) for a reliable result. This demonstrates that it is only galaxies within a 40" radius of any blob that appear to be distinct from the field. 

Thus, when defining the blob region radius for our fiducial analysis (Section \ref{sec:galsample}), we converged on r=40", this being the smallest radius that allows a blob galaxy sample of sufficient size (Section \ref{sec:blobregionradius}), as well as being the maximum size of the \lya\ nebulae's apparent region of influence.

\subsubsection{All Blobs are to Blame}

\begin{figure}
    \centering
    \includegraphics[width=0.5\textwidth]{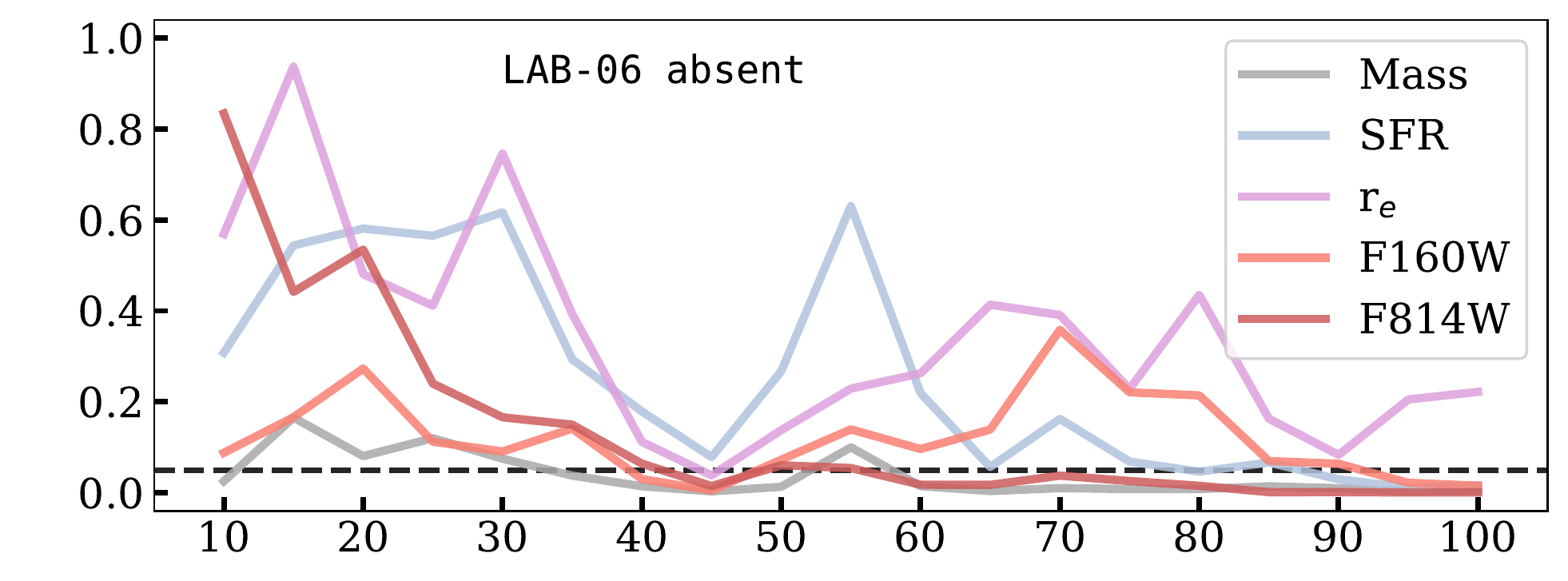}
    \includegraphics[width=0.5\textwidth]{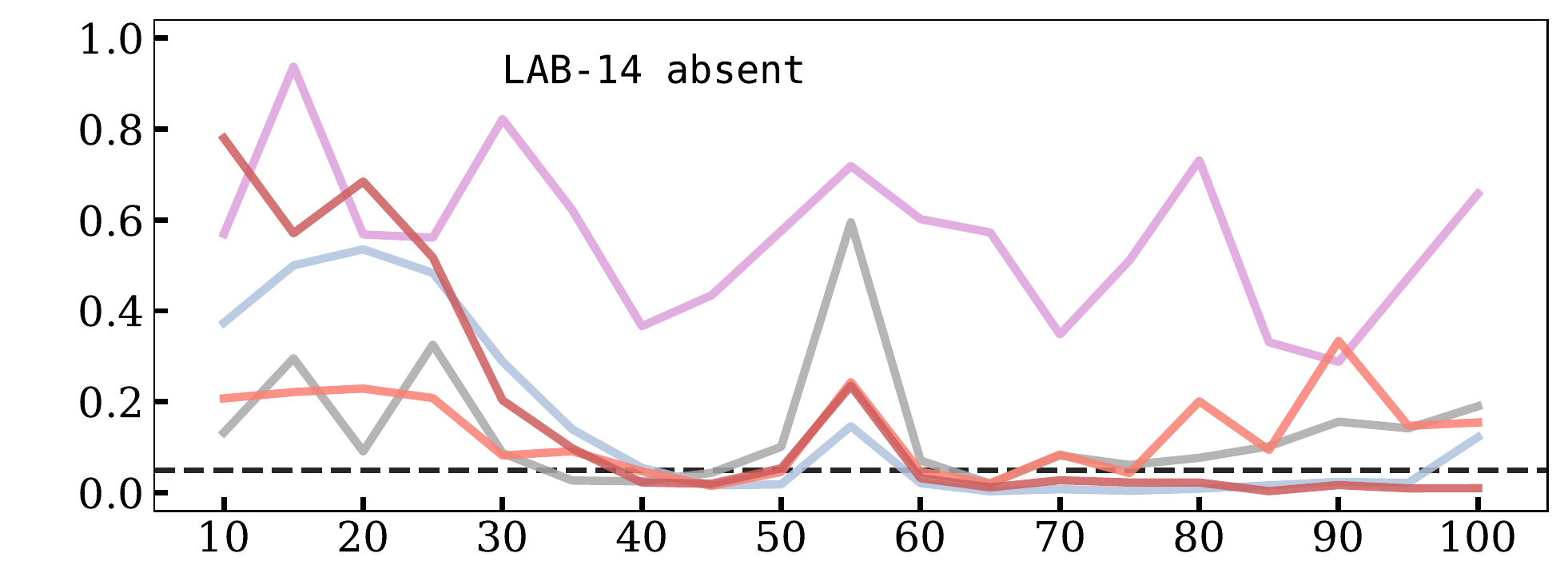}
    \includegraphics[width=0.5\textwidth]{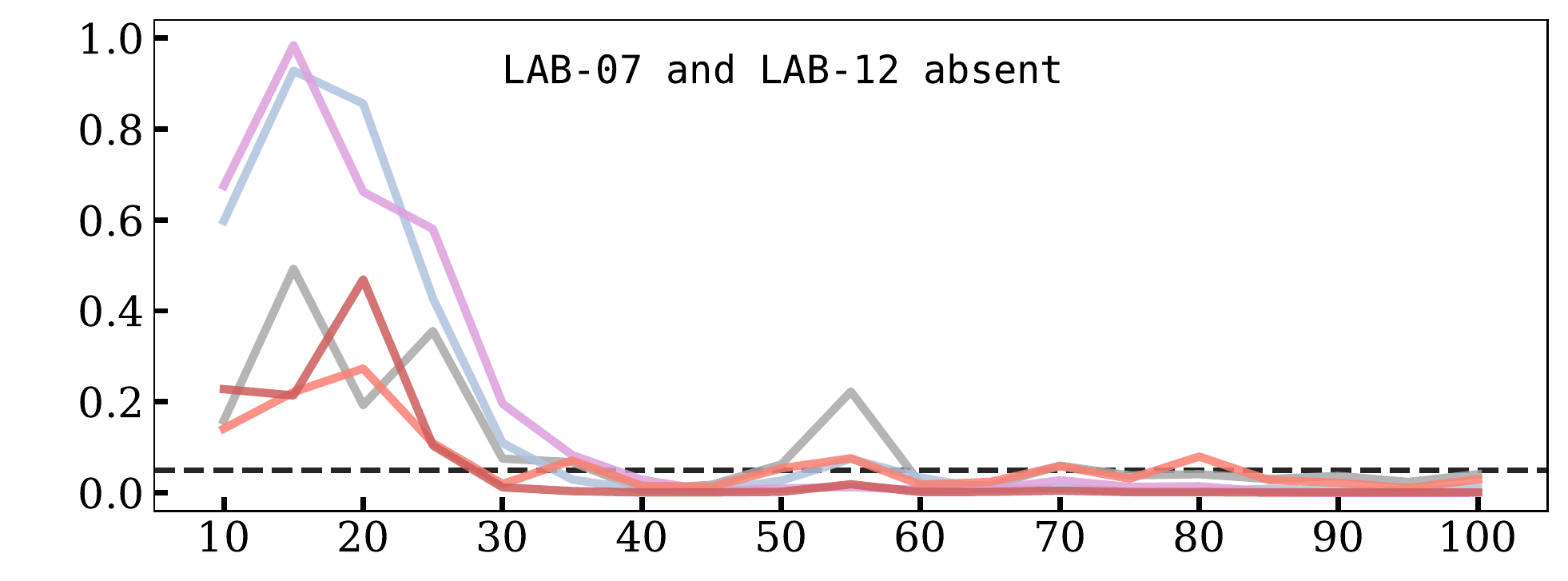}
    \includegraphics[width=0.5\textwidth]{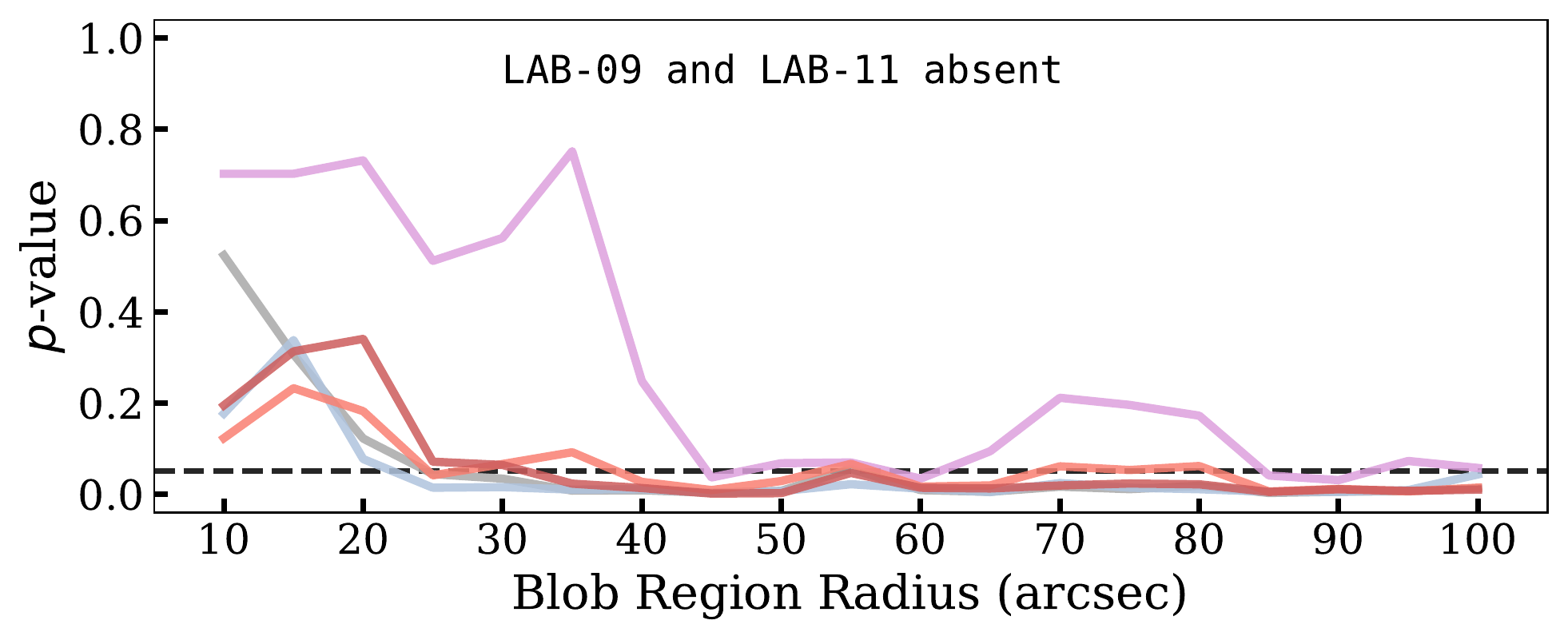}
    \caption{{Removing galaxies in each blob region from the analysis in turn does not eliminate all statistically significant differences between the blob galaxy and field galaxy groups (although LAB-06 and LAB-14 may be contributing more to the signal), indicating that all six blob regions, and not a single special location, are responsible for the observed differences.}}
    \label{fig:GOODSS_pvalcurves_jackknife}
\end{figure}

We considered the possibility that just one or two of the six blob regions contained galaxies sufficiently distinct from the field so as to sway the results, given that we are using a composite blob galaxy sample. Again we allowed the blob region radius around each blob position to grow from 10" to 100", comparing the resulting blob and field galaxies after having removed each blob and blob pair (since two pairs of blobs lie within 55" of each other, see Figure \ref{fig:GOODSS_BlobGals}) from the analysis in turn. Without the contribution of the galaxies associated with one particular blob or another, we still see significant differences between the blob and field galaxies at r$\approx$40", as shown in Figure \ref{fig:GOODSS_pvalcurves_jackknife}, although LAB-06 and LAB-14 may be contributing more to the signal. All of the blob regions, then, appear to contain galaxies that are distinct from the field.

\subsubsection{Random Blob Locations}

As an independent check of our KS test results, we examined the same $p$-value curves over increasing blob region radii when using random \lya\ nebula locations, comparing the resulting groups of mock blob and field galaxy properties in order to see how likely it is that we would see results similar to Figure \ref{fig:GOODSS_pvalcurve} by chance. \citet{Yang2010} surveyed the entire footprint of GOODS-S with 3D-HST coverage, so if random blob-free positions at z=2.3 show galaxy properties that are similarly offset from those in the field, this would cast doubt on our ability to use \lya\ nebulae to identify special areas of influence at this epoch.

\begin{figure}
    \centering
    \includegraphics[width=0.5\textwidth]{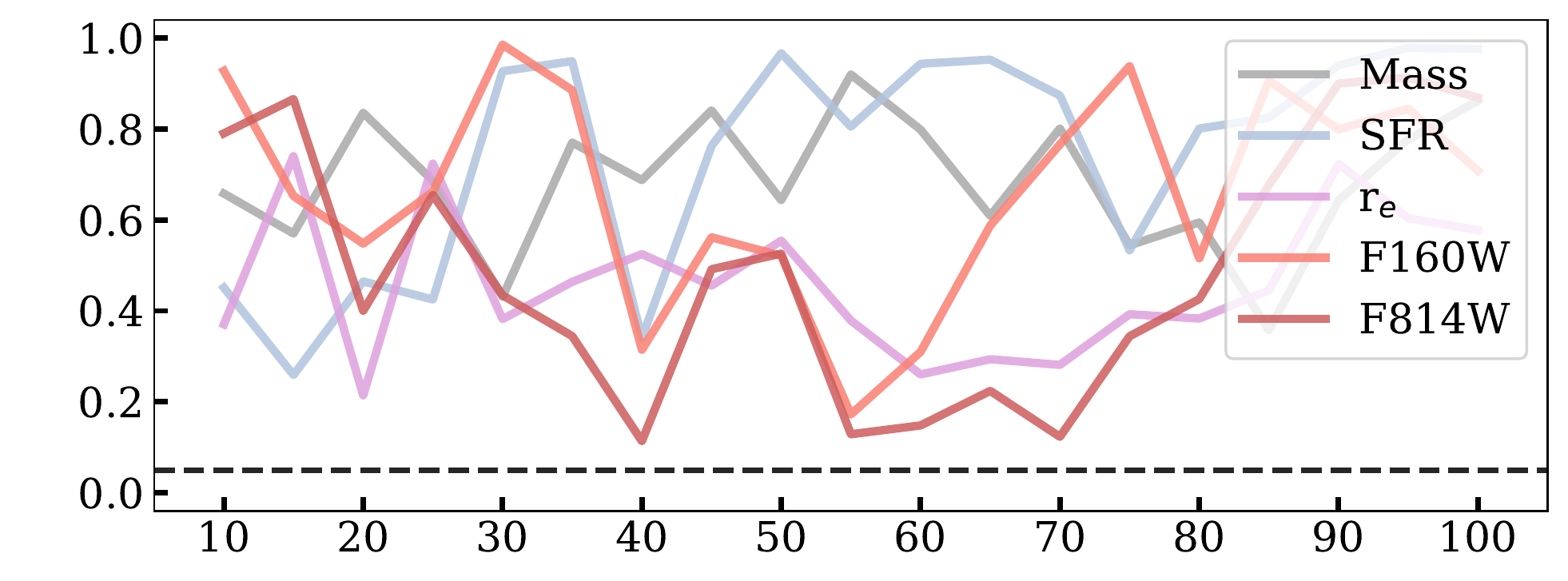}
    \includegraphics[width=0.5\textwidth]{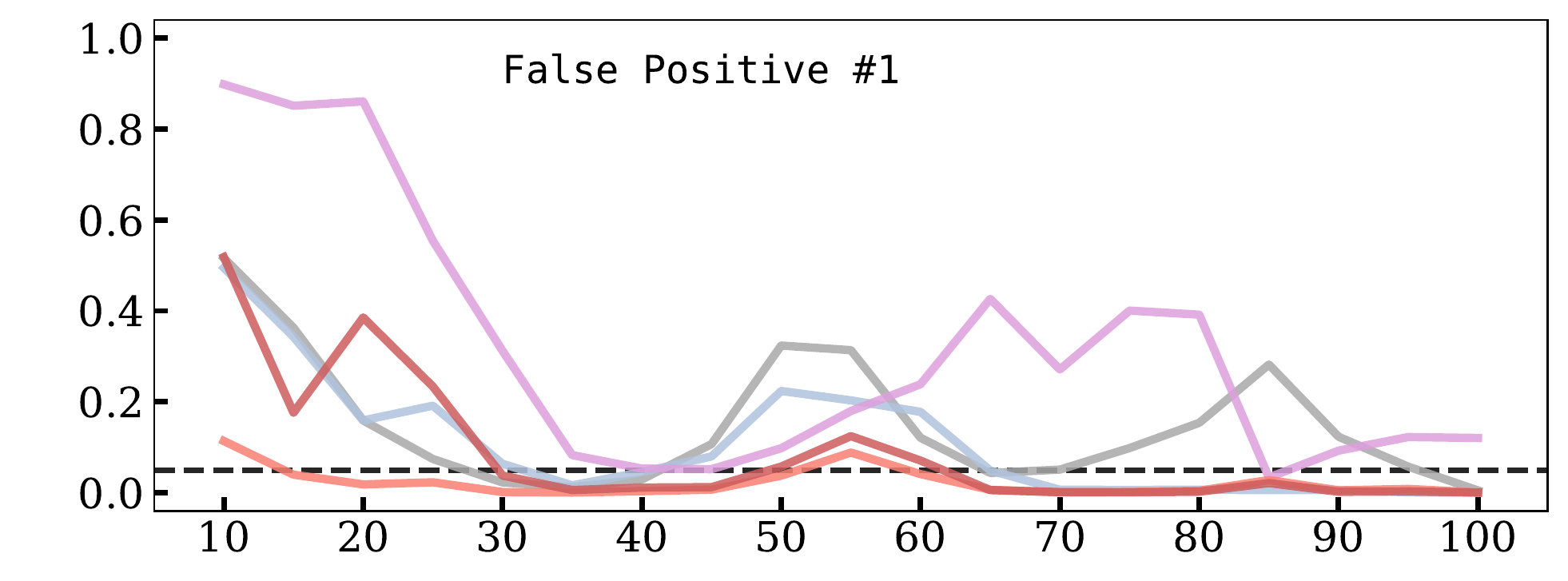}
    \includegraphics[width=0.5\textwidth]{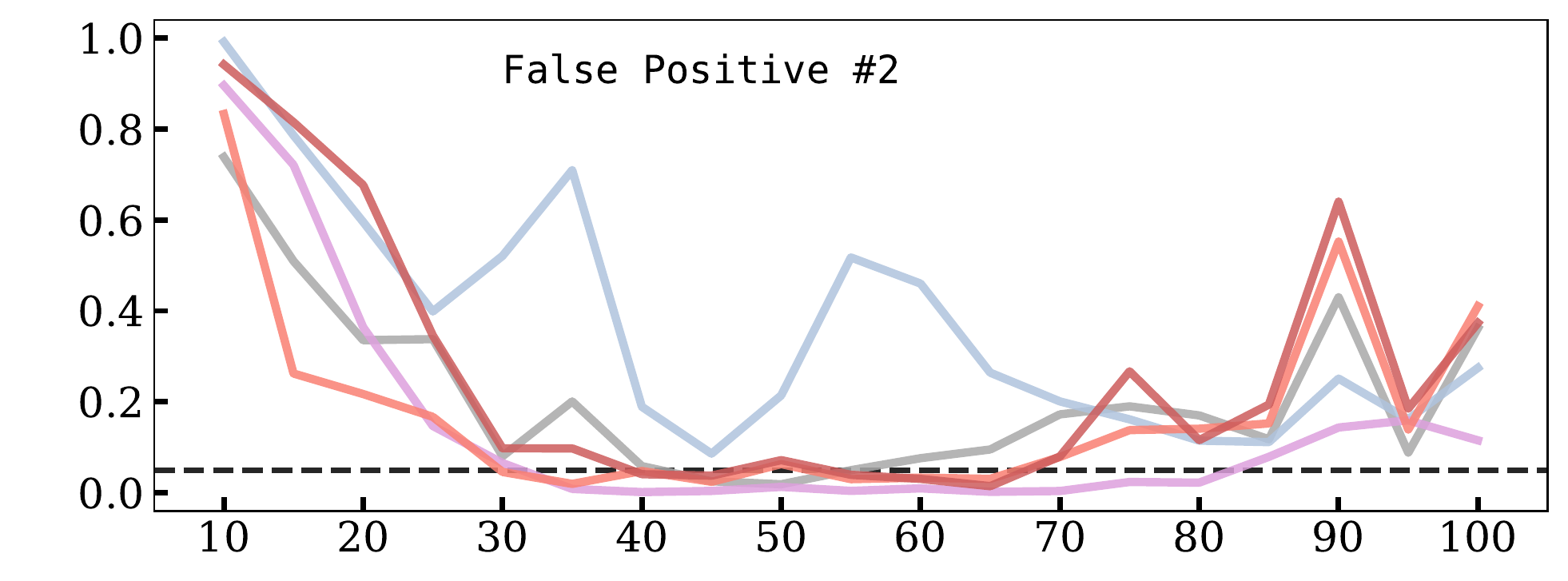}
    \includegraphics[width=0.5\textwidth]{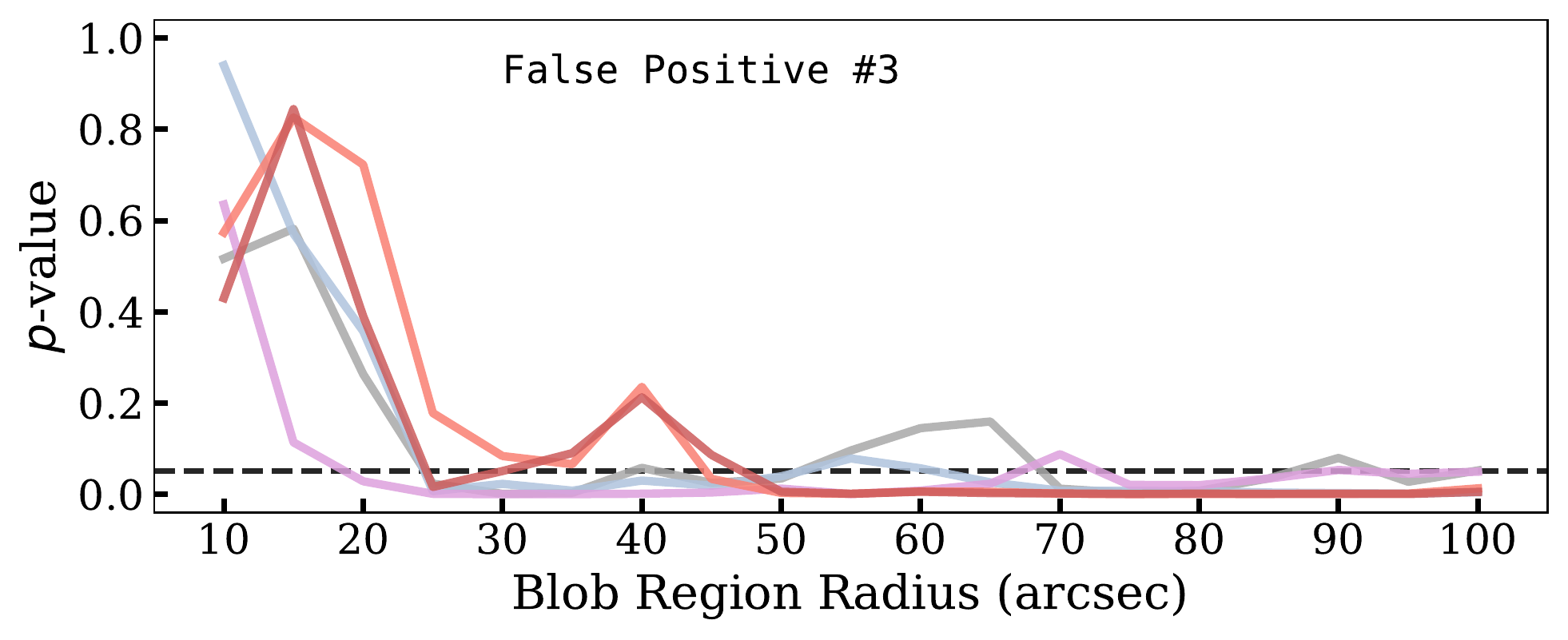}
    \caption{{Examples of KS $p$-value curves from 100 trials using random \lya\ blob locations. The first panel is an example of the typical result. The remaining panels are the three cases which met our ``false positive'' criteria, as discussed in the text, i.e., appearing most similar to our results when using the true \lya\ blob locations (Figure \ref{fig:GOODSS_pvalcurve}).}}
    \label{fig:GOODSS_pvalcurves_100randomlocs}
\end{figure}

To carry out this test, we generated 100 trial sets of 6 random blob positions in GOODS-S, computing the KS test results and p-value curves in each case. The top panel of Figure \ref{fig:GOODSS_pvalcurves_100randomlocs} shows a typical result, with high and meandering $p$-values as a function of blob region radius. By contrast, the real blob locations produced $p$-values that dipped below $p$=0.05 at or before r=40” for 4 out of 5 properties of interest (Figure \ref{fig:GOODSS_pvalcurve}). Out of those 100 trials, only 3 produced $p$-value that meet those same criteria (Figure \ref{fig:GOODSS_pvalcurves_100randomlocs}, \emph{lower three panels}).

The same test was repeated after removing what we may call the ``true'' blob galaxies, those we defined as lying within 40" of any of the 6 actual \lya\ nebula locations with 2.15$\leq$z$\leq$2.45, to ensure that those galaxies were not contaminants in the random location trials. No significant difference was seen in the resulting $p$-value curve plots.

These results correspond to a 3\% false positive rate, consistent with or better than the expected 5\% threshold employed in the KS test. In addition, if we use a stricter definition of significance satisfied by our real data, requiring $p<$0.05 across the range of r=40-50" for 4 out of 5 properties, we find zero false positives out of 100 trials. Thus, the environments around the six true \lya\ nebulae do indeed appear to be ``special'' places in the GOODS-S field at z$\sim$2.3.

\section{Discussion} 
\label{sec:discussion}

\begin{figure}
\centering
\includegraphics[width=0.5\textwidth]{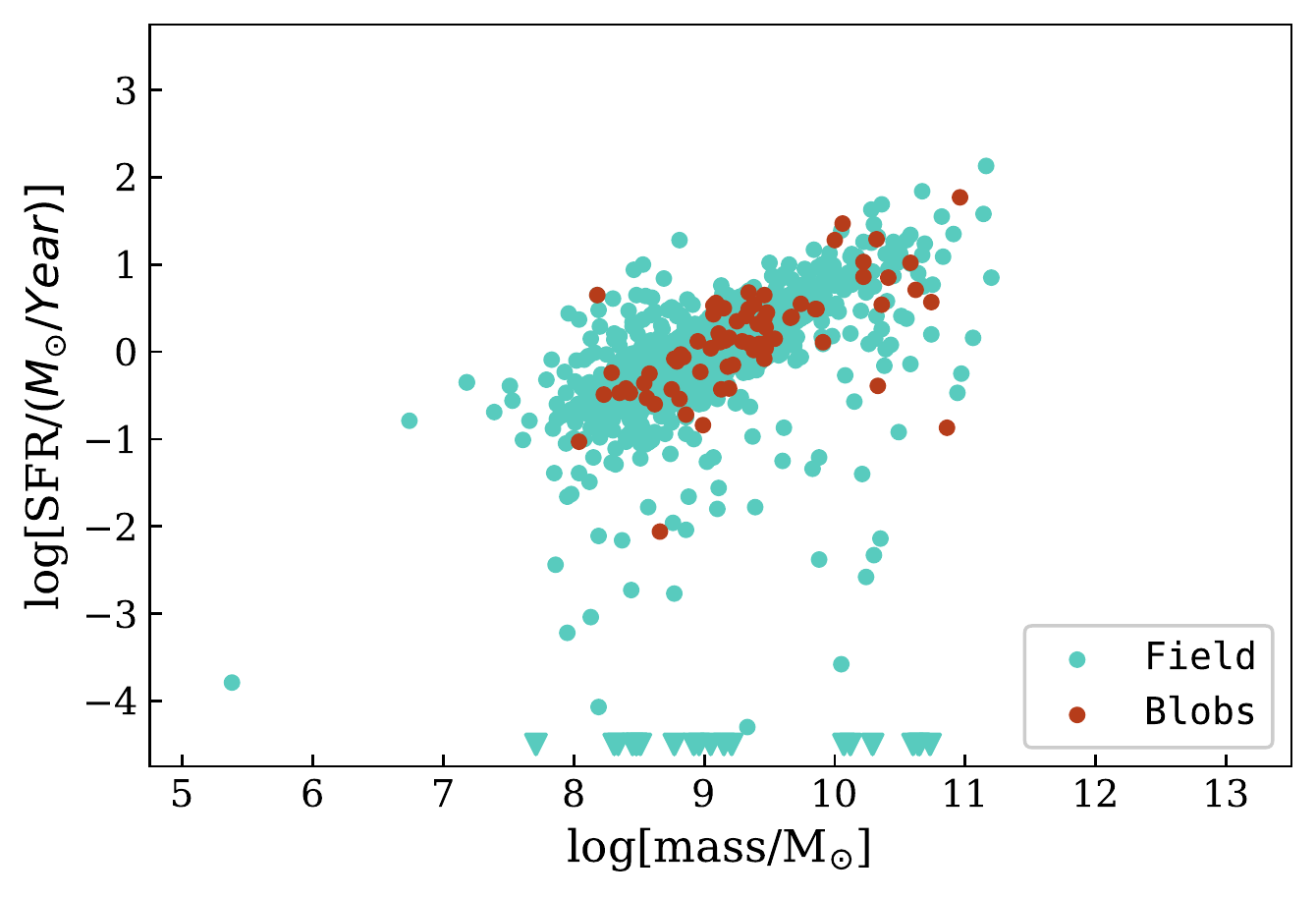}
\caption{SFR versus stellar mass of galaxies associated with blobs ($red$) and those in the field ($blue$). Triangle symbols indicate a non-zero SFR below $10^{-4.5}$ solar masses/year. Both populations appear to follow the same star-forming main sequence, corroborating the result that there is no statistical difference in their specific SFR distributions (Figure \ref{fig:hists_FAST}, \emph{bottom panel}).}
\label{fig:GOODSS_SFR-MassRelation}
\end{figure}

Our investigation has found significant differences between galaxies associated with \lya\ nebulae at z$\sim$2.3 versus those elsewhere in the 170 arcmin${^2}$ GOODS-S field. Under the assumption that Lya nebulae trace overdensities, our results imply that at this redshift, galaxies in denser environments are brighter in F160W and F814W magnitude (n=86, n=74), with higher SFR (n=72) and larger effective radii (n=44). We found that galaxies near \lya\ nebulae populate the same main sequence of star formation (n=72; Figure \ref{fig:hists_FAST}, \emph{bottom}; Figure \ref{fig:GOODSS_SFR-MassRelation}), but that nebula regions have proportionally more high-mass galaxies as well as fewer low-mass ones (n=72; Figure \ref{fig:hists_FAST}, \emph{top}). We note that our results were based on an empirically-derived radius criterion of 40” ($\sim$320 proper kpc at z$\sim$2.3). Assuming these z$\sim$2.3 \lya\ nebulae inhabit galaxy group scale 10$^{13}$ M$\odot$ mass halos \citep{Yang2010}, a standard spherical collapse calculation \citep{Bryan1998} suggests virial radii of 219 proper kpc. Interestingly, this value is within 50\% of the radius where our analysis robustly detects differences in the properties of blob-associated galaxies versus those in the field.

Our findings are in broad agreement with other studies of the dependence of protocluster galaxy properties on environment at high redshift (z=2-4). Many studies have reported that protocluster members skew towards higher stellar masses \citep{Koyama2013, Hatch2011, Steidel2005, Cooke2014, Shimakawa2017, Ito2020}. Protocluster and field galaxies also generally follow the same SFR-stellar mass relation (the so-called star forming main sequence), however here there is somewhat less agreement in the literature. For example, \cite{Steidel2005} examined 72 UV-selected galaxies in a z$\sim$2.3 overdensity and found the mean stellar mass was 2x higher than a comparison group. \cite{Koyama2013} observed 83 H-alpha emitting galaxies (HAEs) in and around a protocluster at z$\sim$2.2, finding that galaxies residing in the densest regions of the cluster appeared to follow the same SFR-stellar mass relation as those in the outskirts, and that the high-density population contained a greater proportion of high-mass galaxies. \cite{Cooke2014} compared HAEs within a 7 arcmin$^2$ overdensity at z$\sim$2.5 to a control group in the field and reported that protocluster and field galaxies showed the same star-forming main sequence with protocluster members skewing towards higher stellar masses, again in good accord with our own results. When considering $>$100 HAEs in another z$\sim$2.5 protocluster and dividing the sample into regions of higher and lower density, \cite{Shimakawa2017} observed the same star-forming main sequence in each but saw a larger proportion of high-mass, high-SFR galaxies in the higher density subset. However, when comparing the lower density group to a superdense subset of the higher density group, \cite{Shimakawa2017} found that the SFRs of the superdense group were boosted for their given stellar mass. Finally, when comparing HAEs within 2 protoclusters at z$\sim$2.2 to a control field sample, \cite{Hatch2011} reported that the protocluster group was 0.8 mag brighter than the control, corresponding to a $\sim$2x greater stellar mass. However, the \cite{Hatch2011} samples did not differ in SFR and the specific SFR of the protocluster population was determined to be lower than that of the field, in contrast to our results. 

At the same time, results on other physical properties are more mixed. For example, our study did not detect a difference in F814W-F160W color, dust extinction (A$_V$), or stellar age, but did find larger effective radii among galaxies in overdensities. \cite{Hatch2011} also observed no significant difference in color when comparing HAEs within 2 protoclusters at z$\sim$2.2 to a control field sample, but by contrast, \cite{Koyama2013} found the HAEs in their high-density group to be redder. \cite{Cooke2014} found a 2x larger median A$_V$ among protocluster members, while \cite{Steidel2005} reported a 2x greater inferred age for the overdensity group. Building on \cite{Steidel2005}, \cite{Peter2007} compared the size distributions of 85 UV-selected star-forming galaxies near the same z$\sim$2.3 protocluster to 63 control galaxies and found no significant difference, yet \cite{Shimakawa2017} found larger sizes for HAEs in the denser regions of the z$\sim$2.5 protocluster. Thus, additional study and larger galaxy samples will be needed to to better understand how galaxy properties correlate with environment at these redshifts.

Taken together, our results add to a suite of high redshift studies consistently finding that galaxies in protoclusters have greater stellar masses than their counterparts in the field. Although varied in their approach to selecting and studying overdense regions, all of these snapshots are consistent with the idea of galaxies in rich environments having matured earlier than their peers in the field, something that is supported by several theoretical studies \citep[][]{Muldrew2015, Chiang2017, Lovell2018}. In this picture, protocluster galaxies virialize and commence star formation sooner, leading them to exhaust their gas reservoirs earlier and evolve into the massive, inactive elliptical galaxies that we observe haunting cluster environments in the local universe. The broad consistency of our results with other complementary protocluster studies lends additional support to the use of \lya\ blobs as markers of dense environments.

\section{Conclusions} 
\label{sec:conclusions}

\lya\ nebulae reside in overdense regions and therefore provide an opportunity to study the impact of such dense environments on the local galaxy population. Using 3D-HST GOODS-S catalog data, we compared traits between galaxies in the vicinity of six \lya\ nebulae and those in the field at z$\sim$2.3. We found statistically significant differences between the two groups in five properties: nebula-associated galaxies are systematically brighter (F160W and F814W) than the field, with correspondingly higher stellar masses, SFRs, and effective radii, but both populations exhibit the same SFR-stellar mass relation. These results are broadly consistent with other studies examining galaxy traits in dense protocluster environments from z=2-4, which have found that the specific star formation rate does not appear to be affected in such regions, but that the fraction of high-mass, high-SFR galaxies is greater in protocluster populations. This suggests that, while star formation proceeds in the same fashion regardless of environment at this epoch, galaxies in dense environments have formed earlier and so are ``ahead'' in their development compared to their peers, a scenario predicted by several theoretical studies. The larger and higher quality galaxy samples expected from the James Webb Space Telescope (JWST) will add clarity to our understanding of high-redshift galaxies and the effects of cluster and group environments on their evolution.

\acknowledgments{The authors would like to thank Kelly Sanderson, Audrey Dijeau, and Daniel Godines Alcantara for helpful discussions. N.K.W. and M.K.M.P. acknowledge support from NSF grant AAG-1813016. This work is based on data obtained for the 3D-HST Treasury Program (GO 12177 and 12328) and the CANDELS Multi-Cycle Treasury Program by the NASA/ESA Hubble Space Telescope, which is operated by the Association of Universities for Research in Astronomy, Incorporated, under NASA contract NAS5-26555.}

\software{Astropy \citep{Astropy, Astropy2018}, SciPy \citep{Virtanen2020}, NumPy \citep{Numpy}, pandas \citep{pandas}, Matplotlib \citep{matplotlib}}

\bibliography{GalaxiesInBlobs}{}

\begin{thebibliography}{}
\expandafter\ifx\csname natexlab\endcsname\relax\def\natexlab#1{#1}\fi
\providecommand{\url}[1]{\href{#1}{#1}}
\providecommand{\dodoi}[1]{doi:~\href{http://doi.org/#1}{\nolinkurl{#1}}}
\providecommand{\doeprint}[1]{\href{http://ascl.net/#1}{\nolinkurl{http://ascl.net/#1}}}
\providecommand{\doarXiv}[1]{\href{https://arxiv.org/abs/#1}{\nolinkurl{https://arxiv.org/abs/#1}}}

\bibitem[{Battaia {et~al.}(2018)Battaia, Chen, Fumagalli, Cai, Rivera, Xu,
  Smail, Prochaska, Yang, Breuck, Battaia, Chen, Fumagalli, Cai, Rivera, Xu,
  Smail, Prochaska, Yang, \& Breuck}]{Battaia2018}
Battaia, F.~A., Chen, C.-C., Fumagalli, M., {et~al.} 2018, A\&A, 620, A202,
  \dodoi{10.1051/0004-6361/201834195}

\bibitem[{Beckwith {et~al.}(2006)Beckwith, Stiavelli, Koekemoer, Caldwell,
  Ferguson, Hook, Lucas, Bergeron, Corbin, Jogee, Panagia, Robberto, Royle,
  Somerville, \& Sosey}]{Beckwith2006}
Beckwith, S. V.~W., Stiavelli, M., Koekemoer, A.~M., {et~al.} 2006, AJ, 132,
  1729, \dodoi{10.1086/507302}

\bibitem[{Brammer {et~al.}(2012)Brammer, van Dokkum, Franx, Fumagalli, Patel,
  Rix, Skelton, Kriek, Nelson, Schmidt, Bezanson, da~Cunha, Erb, Fan,
  Schreiber, Illingworth, Labbé, Leja, Lundgren, Magee, Marchesini, McCarthy,
  Momcheva, Muzzin, Quadri, Steidel, Tal, Wake, Whitaker, \&
  Williams}]{Brammer2012}
Brammer, G.~B., van Dokkum, P.~G., Franx, M., {et~al.} 2012, ApJ Supplement
  Series, 200, \dodoi{10.1088/0067-0049/200/2/13}

\bibitem[{Bryan \& Norman(1998)}]{Bryan1998}
Bryan, G.~L., \& Norman, M.~L. 1998, ApJ, 495, 80

\bibitem[{Cai {et~al.}(2017{\natexlab{a}})Cai, Fan, Yang, Bian, Prochaska,
  Zabludoff, McGreer, Zheng, Green, Cantalupo, Frye, Hamden, Jiang, Kashikawa,
  \& Wang}]{Cai2017a}
Cai, Z., Fan, X., Yang, Y., {et~al.} 2017{\natexlab{a}}, ApJ, 837,
  \dodoi{10.3847/1538-4357/aa5d14}

\bibitem[{Cai {et~al.}(2017{\natexlab{b}})Cai, Fan, Bian, Zabludoff, Yang,
  Prochaska, McGreer, Zheng, Kashikawa, Wang, Frye, Green, Jiang, Cai, Fan,
  Bian, Zabludoff, Yang, Prochaska, McGreer, Zheng, Kashikawa, Wang, Frye,
  Green, \& Jiang}]{Cai2017b}
Cai, Z., Fan, X., Bian, F., {et~al.} 2017{\natexlab{b}}, ApJ, 839,
  \dodoi{10.3847/1538-4357/AA6A1A}

\bibitem[{Chiang {et~al.}(2017)Chiang, Overzier, Gebhardt, \&
  Henriques}]{Chiang2017}
Chiang, Y.-K., Overzier, R.~A., Gebhardt, K., \& Henriques, B. 2017, ApJ
  Letters, 844, \dodoi{10.3847/2041-8213/AA7E7B}

\bibitem[{Collaboration {et~al.}(2013)Collaboration, Robitaille, Tollerud,
  Greenfield, Droettboom, Bray, Aldcroft, Davis, Ginsburg, Price-Whelan,
  Kerzendorf, Conley, Crighton, Barbary, Muna, Ferguson, Grollier, Parikh,
  Nair, Günther, Deil, Woillez, Conseil, Kramer, Turner, Singer, Fox, Weaver,
  Zabalza, Edwards, Bostroem, Burke, Casey, Crawford, Dencheva, Ely, Jenness,
  Labrie, Lim, Pierfederici, Pontzen, Ptak, Refsdal, Servillat, \&
  Streicher}]{Astropy}
Collaboration, A., Robitaille, T.~P., Tollerud, E.~J., {et~al.} 2013, A\&A,
  558, \dodoi{10.1051/0004-6361/201322068}

\bibitem[{Collaboration {et~al.}(2018)Collaboration, Price-Whelan, Sipőcz,
  Günther, Lim, Crawford, Conseil, Shupe, Craig, Dencheva, Ginsburg,
  VanderPlas, Bradley, Pérez-Suárez, de~Val-Borro, Aldcroft, Cruz,
  Robitaille, Tollerud, Ardelean, Babej, Bachetti, Bakanov, Bamford, Barentsen,
  Barmby, Baumbach, Berry, Biscani, Boquien, Bostroem, Bouma, Brammer, Bray,
  Breytenbach, Buddelmeijer, Burke, Calderone, Rodríguez, Cara, Cardoso,
  Cheedella, Copin, Crichton, DÁvella, Deil, Depagne, Dietrich, Donath,
  Droettboom, Earl, Erben, Fabbro, Ferreira, Finethy, Fox, Garrison, Gibbons,
  Goldstein, Gommers, Greco, Greenfield, Groener, Grollier, Hagen, Hirst,
  Homeier, Horton, Hosseinzadeh, Hu, Hunkeler, Ivezić, Jain, Jenness, Kanarek,
  Kendrew, Kern, Kerzendorf, Khvalko, King, Kirkby, Kulkarni, Kumar, Lee, Lenz,
  Littlefair, Ma, Macleod, Mastropietro, McCully, Montagnac, Morris, Mueller,
  Mumford, Muna, Murphy, Nelson, Nguyen, Ninan, Nöthe, Ogaz, Oh, Parejko,
  Parley, Pascual, Patil, Patil, Plunkett, Prochaska, Rastogi, Janga, Sabater,
  Sakurikar, Seifert, Sherbert, Sherwood-Taylor, Shih, Sick, Silbiger,
  Singanamalla, Singer, Sladen, Sooley, Sornarajah, Streicher, Teuben, Thomas,
  Tremblay, Turner, Terrón, van Kerkwijk, de~la Vega, Watkins, Weaver,
  Whitmore, Woillez, \& Zabalza}]{Astropy2018}
Collaboration, A., Price-Whelan, A.~M., Sipőcz, B.~M., {et~al.} 2018, AJ, 156,
  \dodoi{10.3847/1538-3881/aabc4f}

\bibitem[{Cooke {et~al.}(2014)Cooke, Hatch, Muldrew, Rigby, \&
  Kurk}]{Cooke2014}
Cooke, E.~A., Hatch, N.~A., Muldrew, S.~I., Rigby, E.~E., \& Kurk, J.~D. 2014,
  MNRAS, 440, 3262, \dodoi{10.1093/mnras/stu522}

\bibitem[{Dressler(1980)}]{Dressler1980}
Dressler, A. 1980, ApJ, 236, 351, \dodoi{10.1086/157753}

\bibitem[{Gomez {et~al.}(2003)Gomez, Nichol, Miller, Balogh, Goto, Zabludoff,
  Romer, Bernardi, Sheth, Hopkins, Castander, Connolly, Schneider, Brinkmann,
  Lamb, SubbaRao, \& York}]{Gomez2003}
Gomez, P.~L., Nichol, R.~C., Miller, C.~J., {et~al.} 2003, ApJ, 584, 210,
  \dodoi{10.1086/345593}

\bibitem[{Grogin {et~al.}(2011)Grogin, Kocevski, Faber, Ferguson, Koekemoer,
  Riess, Acquaviva, Alexander, Almaini, Ashby, Barden, Bell, Bournaud, Brown,
  Caputi, Casertano, Cassata, Castellano, Challis, Chary, Cheung, Cirasuolo,
  Conselice, Cooray, Croton, Daddi, Dahlen, Davé, de~Mello, Dekel, Dickinson,
  Dolch, Donley, Dunlop, Dutton, Elbaz, Fazio, Filippenko, Finkelstein,
  Fontana, Gardner, Garnavich, Gawiser, Giavalisco, Grazian, Guo, Hathi,
  Hopkins, Huang, Huang, Jha, Kartaltepe, Kirshner, Koo, Lai, Lee, Li, Lotz,
  Lucas, Madau, McCarthy, McGrath, McIntosh, McLure, Mobasher, Moustakas,
  Mozena, Nandra, Newman, Niemi, Noeske, Papovich, Pentericci, Pope, Primack,
  Rajan, Ravindranath, Reddy, Renzini, Rix, Robaina, Rodney, Rosario, Rosati,
  Salimbeni, Scarlata, Siana, Simard, Smidt, Somerville, Spinrad, Straughn,
  Strolger, Telford, Teplitz, Trump, van~der Wel, Villforth, Wechsler, Weiner,
  Wiklind, Wild, Wilson, Wuyts, Yan, \& Yun}]{Grogin2011}
Grogin, N.~A., Kocevski, D.~D., Faber, S.~M., {et~al.} 2011, ApJ Supplement
  Series, 197, \dodoi{10.1088/0067-0049/197/2/35}

\bibitem[{Guo {et~al.}(2013)Guo, Ferguson, Giavalisco, Barro, Willner, Ashby,
  Dahlen, Donley, Faber, Fontana, Galametz, Grazian, Huang, Kocevski,
  Koekemoer, Koo, Mcgrath, Peth, Salvato, Wuyts, Castellano, Cooray, Dickinson,
  Dunlop, Fazio, Gardner, Gawiser, Grogin, Hathi, Hsu, Lee, Lucas, Mobasher,
  Nandra, Newman, \& Wel}]{Guo2013}
Guo, Y., Ferguson, H.~C., Giavalisco, M., {et~al.} 2013, ApJ Supplement Series,
  207, \dodoi{10.1088/0067-0049/207/2/24}

\bibitem[{Harris {et~al.}(2020)Harris, Millman, van~der Walt, Gommers,
  Virtanen, Cournapeau, Wieser, Taylor, Berg, Smith, Kern, Picus, Hoyer, van
  Kerkwijk, Brett, Haldane, del Río, Wiebe, Peterson, Gérard-Marchant,
  Sheppard, Reddy, Weckesser, Abbasi, Gohlke, \& Oliphant}]{Numpy}
Harris, C.~R., Millman, K.~J., van~der Walt, S.~J., {et~al.} 2020, Nature, 585,
  357, \dodoi{10.1038/s41586-020-2649-2}

\bibitem[{Hatch {et~al.}(2011)Hatch, Kurk, Pentericci, Venemans, Kuiper, Miley,
  \& Röttgering}]{Hatch2011}
Hatch, N.~A., Kurk, J.~D., Pentericci, L., {et~al.} 2011, MNRAS, 415, 2993,
  \dodoi{10.1111/J.1365-2966.2011.18735.X}

\bibitem[{Hennawi {et~al.}(2015)Hennawi, Prochaska, Cantalupo, \&
  Arrigoni-Battaia}]{Hennawi2015}
Hennawi, J.~F., Prochaska, J.~X., Cantalupo, S., \& Arrigoni-Battaia, F. 2015,
  Science, 348, 779, \dodoi{10.1126/science.aaa5397}

\bibitem[{Hodges(1958)}]{Hodges1958}
Hodges, J.~L. 1958, Arkiv fü matematik, 3, 469, \dodoi{10.1007/BF02589501}

\bibitem[{Hogg {et~al.}(2004)Hogg, Blanton, Brinchmann, Eisenstein, Schlegel,
  Gunn, McKay, Rix, Bahcall, Brinkmann, \& Meiksin}]{Hogg2004}
Hogg, D.~W., Blanton, M.~R., Brinchmann, J., {et~al.} 2004, ApJ, 601,
  \dodoi{10.1086/381749/FULLTEXT/}

\bibitem[{Hunter(2007)}]{matplotlib}
Hunter, J.~D. 2007, Computing in Science and Engineering, 9, 90,
  \dodoi{10.1109/MCSE.2007.55}

\bibitem[{Ito {et~al.}(2020)Ito, Kashikawa, Toshikawa, Overzier, Kubo,
  Uchiyama, Liang, Onoue, Tanaka, Komiyama, Lee, Lin, Marinello, Martin, \&
  Shibuya}]{Ito2020}
Ito, K., Kashikawa, N., Toshikawa, J., {et~al.} 2020, ApJ, 899,
  \dodoi{10.3847/1538-4357/aba269}

\bibitem[{Kauffmann {et~al.}(2004)Kauffmann, White, Heckman, Ménard,
  Brinchmann, Charlot, Tremonti, \& Brinkmann}]{Kauffmann2004}
Kauffmann, G., White, S.~D., Heckman, T.~M., {et~al.} 2004, MNRAS, 353, 713,
  \dodoi{10.1111/j.1365-2966.2004.08117.x}

\bibitem[{Koekemoer {et~al.}(2013)Koekemoer, Ellis, Mclure, Dunlop, Robertson,
  Ono, Schenker, Ouchi, Bowler, Rogers, Curtis-Lake, Schneider, Charlot, Stark,
  Furlanetto, Cirasuolo, Wild, \& Targett}]{Koekemoer2013}
Koekemoer, A.~M., Ellis, R.~S., Mclure, R.~J., {et~al.} 2013, ApJ Supplement
  Series, 209, \dodoi{10.1088/0067-0049/209/1/3}

\bibitem[{Koyama {et~al.}(2013)Koyama, Kodama, Tadaki, Hayashi, Tanaka, Smail,
  Tanaka, \& Kurk}]{Koyama2013}
Koyama, Y., Kodama, T., Tadaki, K.-I., {et~al.} 2013, MNRAS, 428, 1551,
  \dodoi{10.1093/mnras/sts133}

\bibitem[{Kriek {et~al.}(2009)Kriek, Dokkum, Labbé, Franx, Illingworth,
  Marchesini, \& Quadri}]{Kriek2009}
Kriek, M., Dokkum, P. G.~V., Labbé, I., {et~al.} 2009, ApJ, 700, 221,
  \dodoi{10.1088/0004-637X/700/1/221}

\bibitem[{Lewis {et~al.}(2002)Lewis, Balogh, Propris, Couch, Bower, Offer,
  Bland-Hawthorn, Baldry, Baugh, Bridges, Cannon, Cole, Colless, Collins,
  Cross, Dalton, Driver, Efstathiou, Ellis, Frenk, Glazebrook, Hawkins,
  Jackson, Lahav, Lumsden, Maddox, Madgwick, Norberg, Peacock, Percival,
  Peterson, Sutherland, \& Taylor}]{Lewis2002}
Lewis, I., Balogh, M., Propris, R.~D., {et~al.} 2002, MNRAS, 334, 673,
  \dodoi{10.1046/j.1365-8711.2002.05558.x}

\bibitem[{Lovell {et~al.}(2018)Lovell, Thomas, \& Wilkins}]{Lovell2018}
Lovell, C.~C., Thomas, P.~A., \& Wilkins, S.~M. 2018, MNRAS, 474, 4612,
  \dodoi{10.1093/mnras/stx3090}

\bibitem[{Matsuda {et~al.}(2004)Matsuda, Yamada, Hayashino, Tamura, Yamauchi,
  Ajiki, Fujita, Murayama, Nagao, Ohta, Okamura, Ouchi, Shimasaku, Shioya,
  Taniguchi, Matsuda, Yamada, Hayashino, Tamura, Yamauchi, Ajiki, Fujita,
  Murayama, Nagao, Ohta, Okamura, Ouchi, Shimasaku, Shioya, \&
  Taniguchi}]{Matsuda2004}
Matsuda, Y., Yamada, T., Hayashino, T., {et~al.} 2004, ApJ, 128, 569,
  \dodoi{10.1086/422020}

\bibitem[{McKinney(2010)}]{pandas}
McKinney, W. 2010, 51--56, \dodoi{10.25080/Majora-92bf1922-00a}

\bibitem[{Momcheva {et~al.}(2016)Momcheva, Brammer, van Dokkum, Skelton,
  Whitaker, Nelson, Fumagalli, Maseda, Leja, Franx, Rix, Bezanson, Cunha,
  Dickey, Schreiber, Illingworth, Kriek, Labbé, Lange, Lundgren, Magee,
  Marchesini, Oesch, Pacifici, Patel, Price, Tal, Wake, van~der Wel, \&
  Wuyts}]{Momcheva2016}
Momcheva, I.~G., Brammer, G.~B., van Dokkum, P.~G., {et~al.} 2016, ApJ
  Supplement Series, 225, \dodoi{10.3847/0067-0049/225/2/27}

\bibitem[{Muldrew {et~al.}(2015)Muldrew, Hatch, \& Cooke}]{Muldrew2015}
Muldrew, S.~I., Hatch, N.~A., \& Cooke, E.~A. 2015, MNRAS, 452, 2528,
  \dodoi{10.1093/mnras/stv1449}

\bibitem[{Oemler(1974)}]{Oemler1974}
Oemler, A. 1974, ApJ, 194, 1, \dodoi{10.1086/153216}

\bibitem[{Oke(1974)}]{Oke1974}
Oke, J.~B. 1974, ApJ Supplement Series, 27, 21, \dodoi{10.1086/190287}

\bibitem[{Peng {et~al.}(2010)Peng, Lilly, Kovač, Bolzonella, Pozzetti,
  Renzini, Zamorani, Ilbert, Knobel, Iovino, Maier, Cucciati, Tasca, Carollo,
  Silverman, Kampczyk, Ravel, Sanders, Scoville, Contini, Mainieri, Scodeggio,
  Kneib, Evre, Bardelli, Bongiorno, Caputi, Coppa, Torre, Franzetti, Garilli,
  Lamareille, Borgne, Brun, Mignoli, Montero, Ricciardelli, Tanaka, Tresse,
  Vergani, Welikala, Zucca, Oesch, Abbas, Barnes, Bordoloi, Bottini, Cappi,
  Cassata, Cimatti, Fumana, Hasinger, Koekemoer, Leauthaud, Maccagni, Marinoni,
  Mccracken, Memeo, Meneux, Nair, Porciani, Presotto, \& Scaramella}]{Peng2010}
Peng, Y.-J., Lilly, S.~J., Kovač, K., {et~al.} 2010, ApJ, 721,
  \dodoi{10.1088/0004-637X/721/1/193}

\bibitem[{Peter {et~al.}(2007)Peter, Shapley, Law, Steidel, Erb, Reddy, \&
  Pettini}]{Peter2007}
Peter, A. H.~G., Shapley, A.~E., Law, D.~R., {et~al.} 2007, ApJ, 668, 23,
  \dodoi{10.1086/521184}

\bibitem[{Prescott {et~al.}(2008)Prescott, Kashikawa, Dey, Matsuda, Prescott,
  Kashikawa, Dey, \& Matsuda}]{Prescott2008}
Prescott, M. K.~M., Kashikawa, N., Dey, A., {et~al.} 2008, ApJ Letters, 678,
  L77, \dodoi{10.1086/588606}

\bibitem[{Razali \& Wah(2011)}]{Razali2011}
Razali, N.~M., \& Wah, Y.~B. 2011, Journal of Statistical Modeling and
  Analytics, 2, 21

\bibitem[{Saito {et~al.}(2006)Saito, Shimasaku, Okamura, Ouchi, Akiyama, \&
  Yoshida}]{Saito2006}
Saito, T., Shimasaku, K., Okamura, S., {et~al.} 2006, ApJ, 648, 54,
  \dodoi{10.1086/505678}

\bibitem[{Scholz \& Stephens(1987)}]{Scholz1987}
Scholz, F.~W., \& Stephens, M.~A. 1987, Journal of the American Statistical
  Association, 82, 918, \dodoi{10.2307/2288805}

\bibitem[{Shimakawa {et~al.}(2018)Shimakawa, Kodama, Hayashi, Prochaska,
  Tanaka, Cai, Suzuki, Tadaki, \& Koyama}]{Shimakawa2017}
Shimakawa, R., Kodama, T., Hayashi, M., {et~al.} 2018, MNRAS, 473, 1977,
  \dodoi{10.1093/mnras/stx2494}

\bibitem[{Skelton {et~al.}(2014)Skelton, Whitaker, Momcheva, Brammer, van
  Dokkum, Labbé, Franx, van~der Wel, Bezanson, Cunha, Fumagalli, Schreiber,
  Kriek, Leja, Lundgren, Magee, Marchesini, Maseda, Nelson, Oesch, Pacifici,
  Patel, Price, Rix, Tal, Wake, \& Wuyts}]{Skelton2014}
Skelton, R.~E., Whitaker, K.~E., Momcheva, I.~G., {et~al.} 2014, ApJ Supplement
  Series, 214, \dodoi{10.1088/0067-0049/214/2/24}

\bibitem[{Steidel {et~al.}(2005)Steidel, Adelberger, Shapley, Erb, Reddy, \&
  Pettini}]{Steidel2005}
Steidel, C.~C., Adelberger, K.~L., Shapley, A.~E., {et~al.} 2005, ApJ, 626, 44,
  \dodoi{10.1086/429989}

\bibitem[{Steidel {et~al.}(2000)Steidel, Adelberger, Shapley, Pettini,
  Dickinson, \& Giavalisco}]{Steidel2000}
---. 2000, ApJ, 532, 170, \dodoi{10.1086/308568}

\bibitem[{van~der Wel {et~al.}(2012)van~der Wel, Bell, Häussler, McGrath,
  Chang, Guo, McIntosh, Rix, Barden, Cheung, Faber, Ferguson, Galametz, Grogin,
  Hartley, Kartaltepe, Kocevski, Koekemoer, Lotz, Mozena, Peth, \&
  Peng}]{vanderwel2012}
van~der Wel, A., Bell, E.~F., Häussler, B., {et~al.} 2012, ApJ Supplement
  Series, 203, \dodoi{10.1088/0067-0049/203/2/24}

\bibitem[{Virtanen {et~al.}(2020)Virtanen, Gommers, Oliphant, Haberland, Reddy,
  Cournapeau, Burovski, Peterson, Weckesser, Bright, van~der Walt, Brett,
  Wilson, Millman, Mayorov, Nelson, Jones, Kern, Larson, Carey, İlhan Polat,
  Feng, Moore, VanderPlas, Laxalde, Perktold, Cimrman, Henriksen, Quintero,
  Harris, Archibald, Ribeiro, Pedregosa, \& van Mulbregt}]{Virtanen2020}
Virtanen, P., Gommers, R., Oliphant, T.~E., {et~al.} 2020, Nature Methods, 17,
  261, \dodoi{10.1038/s41592-019-0686-2}

\bibitem[{Yang {et~al.}(2010)Yang, Zabludoff, Eisenstein, \& Davé}]{Yang2010}
Yang, Y., Zabludoff, A., Eisenstein, D., \& Davé, R. 2010, ApJ, 719, 1654,
  \dodoi{10.1088/0004-637X/719/2/1654}

\bibitem[{Yang {et~al.}(2009)Yang, Zabludoff, Tremonti, Eisenstein, \&
  Davé}]{Yang2009}
Yang, Y., Zabludoff, A., Tremonti, C., Eisenstein, D., \& Davé, R. 2009, ApJ,
  693, 1579, \dodoi{10.1088/0004-637X/693/2/1579}

\end{thebibliography}
\bibliographystyle{aasjournal}

\end{document}